\documentclass[preprint]{aastex}

\received{}
\accepted{}

\slugcomment{Draft, to appear in ApJ}

\shortauthors{Tran et al.}
\shorttitle{ISOCAM spectroscopy of ULIRGs}

\begin{document}

\title{ISOCAM-CVF 5-12$\mu$m spectroscopy of Ultraluminous Infrared 
Galaxies
   \footnote{Based on observations with ISO, an ESA project with
   instruments funded by ESA Member States (especially the PI
   countries: France, Germany, the Netherlands and the United Kingdom)
   with the participation of ISAS and NASA} }

\author{Q.D. Tran\altaffilmark{1},
  D. Lutz\altaffilmark{1},
  R. Genzel\altaffilmark{1},
  D. Rigopoulou\altaffilmark{1},
  H.W.W. Spoon\altaffilmark{2},
  E. Sturm\altaffilmark{1},
  M. Gerin\altaffilmark{3},
  D.C. Hines\altaffilmark{4},
  A.F.M. Moorwood\altaffilmark{2},
  D.B. Sanders\altaffilmark{5},
  N. Scoville\altaffilmark{6},
  Y. Taniguchi\altaffilmark{7},
  M. Ward\altaffilmark{8}}
 
\altaffiltext{1}{Max-Planck-Institut f\"ur extraterrestrische 
Physik, Postfach 1603, 85740 Garching, Germany}
\altaffiltext{2}{European Southern Observatory,
Karl-Schwarzschild-Stra\ss\/e 2, 85748 Garching, Germany}
\altaffiltext{3}{Ecole Normale Superieure - Radioastronomie, 24 Rue Lhomond,
F-75231 Paris, France}
\altaffiltext{4}{Steward Observatory, The University of Arizona
933 N. Cherry Ave., Tucson, AZ 85721, USA}
\altaffiltext{5}{Institute for Astronomy, University of Hawaii, 2680
Woodlawn Drive, Honolulu, HI 96822, USA}
\altaffiltext{6}{California Institute of Technology, Pasadena, CA 91125,USA}
\altaffiltext{7}{Astronomical Institute, Tohoku University,
Aramaki, Aoba, Sendai 980-8758, Japan}
\altaffiltext{8}{Department of Physics and Astronomy, Leicester 
University, Leicester, LE1 7RH, UK}

\email{dantran@mpe.mpg.de}

\begin{abstract}

We present low resolution mid infrared spectra of 16 ultraluminous
infrared galaxies (ULIRGs) obtained with the CVF spectroscopy mode of
ISOCAM on board the Infrared Space Observatory ISO. Our sample
completes previous ISO spectroscopy of ultra- and hyperluminous
infrared galaxies towards higher luminosities. The combined samples
cover an infrared luminosity range of $\sim10^{12 -13.1} L_{\sun}$.

To discriminate AGN and starburst activity, we use the AGN-related
mid-infrared continuum and the starburst-related 6.2, 7.7, 8.6,
11.3$\mu$m mid infrared emission bands attributed to aromatic
carbonaceous material. For about half of the high luminosity ULIRGs
studied here, strong aromatic emission bands suggest starburst
dominance. Other spectra are dominated by a strong AGN-related
continuum with weak superposed emission features of uncertain nature.
Our sample contains one unusual example, IRAS F00183-7111, of an AGN
that is highly obscured even in the mid-infrared.
 
An improved method to quantitatively characterize the relative
contribution of star formation and AGN activity to the mid-infrared
emission of ULIRGs is presented. The ULIRG spectra are fitted by a
superposition of a starburst and an AGN spectrum, both of which may be
obscured at different levels. Models in which starburst and AGN
obscuration differ are significantly more successful than models with
a single extinction.  Previous results based on a simpler
line-to-continuum measure of aromatic emission strength are confirmed,
further supporting the robustness of the aromatic emission feature as
a diagnostic of ULIRG power sources.

As dominant source of the bolometric luminosity, starbursts prevail at
the lower end and AGNs at the higher end of this range. The transition
between mostly starburst and mostly AGN powered occurs at $\sim
10^{12.4}$ to $10^{12.5} L_{\sun}$, and individual luminous
starbursts are found up to $\sim 10^{12.65} L_{\sun}$.

\end{abstract}


\section{Introduction}

Recently, mid infrared spectroscopy has been used increasingly as a
tool for studies of ultraluminous infrared galaxies (ULIRGs,
$L_{\rm IR}>10^{12}L_{\sun}$, see Sanders \& Mirabel (1996) for a
review). The presence of starbursts and active galactic nuclei (AGN)
in ULIRGs (and coexistence in some of them) has been known for quite
some time, but the question which of the two dominates the luminosity
has been difficult to answer because of the large columns of obscuring
dust found towards the nuclear regions of these gas- and dust-rich
systems.  High sensitivity and complete coverage of the infrared
spectrum by ESA's Infrared Space Observatory (ISO, Kessler et
al. 1996) have considerably advanced the use of infrared spectroscopy
to penetrate this obscuring dust, and to probe for the sources of the
huge luminosity of ULIRGs.

Fine structure line and aromatic emission feature observations with
ISO-SWS and ISOPHOT-S of a sample of 15 bright ULIRGs suggest that
most ULIRGs are predominantly starburst-powered \citep{genzel98}. The
PAH method has been extended to a larger sample
\citep{lutz98,rigopoulou99}, allowing to probe for evolutionary
effects expected in a scenario where starburst activity gives way to a
quasar-like AGN buried inside the ULIRG \citep{sanders88}. Comparison
of optical and mid-infrared spectroscopic diagnostics
\citep{lutz99,taniguchi99} demonstrated surprisingly good agreement if
optical LINERs are interpreted as starbursts. This finding also
suggests that AGN in ULIRGs usually make their presence known
optically at least in certain directions, instead of being fully
embedded by large obscuring columns of dust.

This paper presents the result of a solicited program in ISO open
time, which addresses two main issues only partly covered by previous
work.  First, and most important, at which luminosity does the
transition from `predominantly starburst-powered' to `predominantly
AGN-powered' occur?  On one hand, the ISO observations of Genzel et
al. (1998) and Lutz et al. (1998) suggest that not only luminous
infrared galaxies with $L_{\rm IR}<10^{12}L_{\sun}$, but also most ULIRGs
above this threshold are predominantly starburst powered. On the other
hand, optical and mid-infrared observations of hyperluminous infrared
galaxies
\citep{hines95,hines99,taniguchi97,aussel98} support a previous 
consensus that AGN dominate these hyperluminous systems (but note the
intriguing possibility that far-infrared peaks in some luminous AGN
may trace even higher star formation related luminosities, e.g. Haas
et al. 1998). The luminosity at which {\em local} ULIRGs ($z < 0.4$) --
on average -- switch from starburst to AGN dominated is a quantity
with implications for various fields ranging from the interpretation
of sources found in recent submm surveys
\citep{hughes98,barger98} to explanation of the hard X-ray background
\citep{comastri95}. Simple arguments based on the free fall time of
the gas mass concentrated in the inner region of ULIRGs and the
nucleosynthesis efficiency suggest a maximum luminosity of a starburst
event approaching $10^{13}L_{\sun}$ \citep{heckman94}, the limit
depending on gas mass, spatial scale, and initial mass function
properties. Observational evidence is clearly needed.  Despite its
size of $\sim$60 sources, the ULIRG sample presented by Lutz et
al. (1998) and Rigopoulou et al. (1999) is not optimal for this task:
It is selected basically as flux limited at 60$\mu$m and dominated by
low to moderate luminosity ULIRGs, because of the steep ULIRG
luminosity function \citep{sanders96}. The preliminary analysis of
Lutz et al. (1998) in the mid-infrared and a recent study of Veilleux
et al. (1999) in the optical/NIR infrared showed that the presence of
AGN in ULIRGs increases with the luminosity of the object.  However,
these studies clearly call for improved statistics above $\sim
10^{12.3} L_{\sun}$ in order to better determine the transition to
AGN-dominated systems.

The second issue addressed by this paper is related to the use of PAH
features as an AGN/starburst diagnostic. Mid-infrared spectra of most
galaxies show the 6.2, 7.7, 8.6, and 11.3$\mu$m features attributed to
aromatic carbonaceous material
\citep{duley81,leger84,sakata87,papoular89}. Among several popular
designations for these bands will be designated hereafter by the terms
PAH (polycylic aromatic hydrocarbon) or UIB (unidentified infrared
bands, because of remaining uncertainties on the precise nature of the
aromatic carrier). Groundbased observations of these features and a
companion at 3.3$\mu$m first demonstrated that their equivalent width
is larger in starburst galaxies than in classical AGNs
\citep{moorwood86,roche91}. ISO spectroscopy has further
strengthened this link by demonstrating the anti-correlation between
feature strength relative to the continuum and the ionization state of
the gas \citep{genzel98}. Spatially resolved ISOCAM-CVF observations
of nearby AGN support this interpretation by showing the PAH features
contrast to be weak near the central AGN but strong in the
circumnuclear region likely dominated by star formation (e.g. Cen A,
NGC 1068: \cite{laurent99}, Circinus: \cite{moorwood99}).  Despite this 
solid empirical basis, special care is required in analyzing PAH spectra of
ULIRGs for AGN or starburst dominance.  All components of the spectrum
-- continuum and emission features, AGN or starburst -- may be
unusually obscured, and the limited wavelength coverage of ISOPHOT-SL
spectra (5.8-11.6$\mu$m observed wavelength) makes continuum
definition difficult. Here, the extended wavelength range of
ISOCAM-CVF (5-16.5$\mu$m) is beneficial in reaching the long
wavelength side of the broad 9.7$\mu$m silicate absorption feature.

Our paper is organized as follows: We discuss sample selection,
observations and data analysis in Section 2. Section 3 presents the
observational results. Section 4 establishes a new quantitative
method to determine the contribution of starburst and AGN activity to
the mid-infrared spectra, and compares it with the method used by
Genzel et al. (1998), Lutz et al. (1998) and Rigopoulou et
al. (1999). In Section 5, we discuss the properties of ULIRGs as a
function of luminosity, combining our sample with the results of the
ISOPHOT-S sample and with ISOCAM results on HYLIRGs. Finally, we
conclude in section 6.

\section{Observations and data reduction}

\subsection{Sample and observing strategy}

The solicited observing program presented here was defined in
the last third of the ISO mission when it had become apparent that low
resolution spectroscopy of ULIRGs was a promising tool, and that
studies of ULIRGs needed to be extended towards higher
luminosities. The sample selection is strongly driven by the
visibility constraints of the ISO satellite during the final six
months of the mission. ULIRGs were selected from the samples of
Fisher et al. (1995), Kim (1995), Clements et al. (1996), and the QDOT
sample (Lawrence et al. 1999). Six sources were chosen from these
catalogues with $L_{\rm IR}>10^{12.5}L_{\sun}$ that had reasonable ISO
visibility during the remaining mission, plus an additional ten
slightly lower luminosity sources that could be accommodated within
the available observing time. Table~\ref{tab:sources} lists basic
properties of the sample galaxies. Throughout this paper, we follow
the convention reported by Sanders \& Mirabel (1996) to compute the
8-1000$\mu$m luminosity $L_{\rm IR}$, adopting
H$_0$=75\,km\,s$^{-1}$\,Mpc$^{-1}$ and q$_0$=0.5, and IRAS FSC fluxes
( IRAS 03521+0028, IRAS 18030+0705: PSC). Most of our sources are
lacking detections in some of the IRAS photometric bands.  In those
cases, we replaced the missing fluxes by estimates based on the
60$\mu$m fluxes and average colors of the Sanders et al. (1988)
ULIRGs. The observations included the source IRAS 03452+2320 taken
from the sample of Fisher et al. (1995), which is not listed in
Table~\ref{tab:sources}.  We did not detect a signal at its
position. We believe this reflects an infrared cirrus related
misidentification in the Fisher et al. (1995) sample rather than
mid-infrared faintness of a true ULIRG, since Crawford et al. (1996)
failed to detect the same source in radio continuum.

Our sample is certainly heterogeneous : it was selected from ULIRG
surveys of different limiting 60$\mu$m flux, with a preference for
high luminosity targets but not implementing a strict lower cutoff,
and strongly driven by ISO visibilities.  None of the selection
criteria, however, should significantly bias our sample in mid-IR
spectral properties or AGN content. In particular, no AGN-related IRAS
color criteria like the 25/60$\mu$m flux ratio were applied 
\citep{degrijp85,sanders88a}. A slight
tendency towards favoring AGN at higher redshifts/luminosities is
induced by the 60$\mu$m selection which is effectively at
$\sim$45$\mu$m rest wavelength for the most distant targets. We have
not attempted to correct for this since detailed far-infrared spectral
energy distributions are unavailable for these faint IRAS sources, and
since the effect will be negligible for our small and still fairly
local sample. The sample is hence suited to study the AGN content of
ULIRGs, bridging the gap between the lower luminosity ULIRGs studied
with ISOPHOT-S \citep{lutz98,rigopoulou99} and the two HyperLuminous
Infrared Galaxies (HYLIRGs) observed by ISO
\cite{taniguchi97,aussel98}. Figure~\ref{Dist_Lir} shows the
distribution of infrared luminosities both for our sample and the
combined ISO database.

The observations presented here used the Circular Variable
Filter (CVF) mode of the ISOCAM camera \citep{cesarsky96} rather than
the ISOPHOT-S mode with which most other low resolution ULIRG spectra
were obtained and which is efficient in building a large sample of
relatively bright sources. The ultimate sensitivity achieved in a long
CVF integration time is better, and the need for stabilization
overheads was no major concern for our moderate number of very faint
sources.  In addition, the extended long wavelength coverage is
crucial in targeting objects up to z$\sim$0.4, and in
probing an extended rest wavelength range.  Since our sources are
expected to be spatially unresolved, we have chosen the most sensitive
setting with 6\arcsec\/ pixels and a total FOV of about
3\arcmin\/. The observed wavelength range varies from source to
source, the short end being defined by 5$\mu$m rest wavelength and the
long end either by 12$\mu$m rest wavelength or, for high redshift
sources, by 16$\mu$m observed wavelength where the sensitivity of CVF
observations starts to decrease quickly. In order to increase
redundancy and to enable correction of the transients of the ISOCAM
detector, each CVF scan has been done in up and down direction.  To
reconcile the need for detector stabilization (10 frames of 5sec
exposure at each wavelength) with the total number of wavelength steps
and practical limits on observing time, we have executed up- and
down-scans with interleaved wavelength steps. The total observing time
per source including overheads was typically 2.5 hours. Each final
spectrum is composed by about 120 wavelength steps and has an average
resolution of $R \approx 35$.

\subsection{Data reduction}  

Reductions have been carried out with the Cam Interactive Analysis
(CIA) software\footnote{CIA is a joint development by ESA Astrophysics
division and the ISOCAM consortium led by the ISOCAM PI, C. Cesarsky,
Direction des Sciences de la Mati\`ere, CEA, France.}. The dark
current has been estimated with the dark model described by Biviano \&
Sauvage (1999). This gives better results than using the two offset
observations we have taken before and after the CVF observation.  We
used the multi-median resolution deglitching method \citep{starck99}
to remove glitches, and then additionally deglitched
manually a square of 3 by 3 pixels centered on the source. To correct
the transient effect of the detector the inversion method developed at 
IAS \citep{coulais99} was used. The calibration of the flux was
performed using the standard spectral response function of CIA.

Appendix~\ref{PSF_red} presents a reduction scheme which allows
to quantify the spectrum of a target, with its S/N ratio and the
systematic error introduced by the reduction method (the total
uncertainty on the flux is about 5~mJy). It also allows us to
determine the position of the fairly faint source with sub-pixel
accuracy. The typical uncertainty of the position is 5\arcsec,
including S/N and ISO pointing uncertainty.

One target (IRAS 00406-3127) of the set had not been acquired by
ISOCAM because of a telemetry drop. For IRAS F02115+0226 there was no
source detection.  The S/N ratio observed for IRAS F02455-2220 is not
high enough to reliably determine the PSF
correction described in the Appendix. However, the observed spectrum
can be taken as a lower limit, using only the signal from the brightest
pixel. For the remaining 13 targets, the data reduction has been
applied successfully.

\section{Results \label{results}}

The rest-frame spectra for the 13 sources are shown in
Figure~\ref{results_spectrum}. The shaded confidence area around the
spectrum represents the RMS plus the systematic error.
Table~\ref{tab:sources} lists the position of the sources as obtained
by the PSF fitting method (see Appendix), and the IRAS
fluxes. All targets were found to be point like sources at ISOCAM
resolution. One target has been found to be significantly offset from
the IRAS PSC position: IRAS 18030+0705. The offset is about one
arcminute. We checked the original IRAS data using XSCANPI and found a
source with fluxes similar to those listed in the PSC, but centered on
the ISOCAM position. The origin of this position discrepancy is
unknown.  The FSC fluxes were adopted for the ULIRGs detected by ISOCAM
except for IRAS 18030+0705 and IRAS 03521+0028 for which we used PSC
fluxes.

>From previous studies, the following components are expected to be
present in ULIRG spectra, but not necessarily together in {\it one}
spectrum :
\begin{itemize}
\item [-] UIB emission features at 6.2, 7.7, 8.6 and 11.3 microns, mainly 
originating from star forming regions of the galaxies. This UIB
emission can be also coupled with a continuum of hot dust present in
the most energetic \ion{H}{2} regions of the starburst.
\item[-] An AGN continuum from the hot dust present in the vicinity of the
  AGN, e.g. in a dust torus or at larger scale in the narrow line
region.
\item[-] Strong absorption, both continuous and in features like the 
silicate 9.7 $\mu$m one. Absorption may arise either in the surroundings of
the AGN or the larger scale ISM. The amount of absorption may be
different for AGN and for starburst.
\end{itemize}
The silicate feature at 9.7$\mu$m that is usually found in absorption
may also mimic emission features.  In dust configurations as in some
torus models \citep{pier92}, self-absorbed silicate emission can
create an emission peak near 8$\mu$m that is superficially similar to
a faint UIB emission and may be observed in one HYLIRG
\citep{taniguchi97}. Discrimination of such a `fake UIB' peak and
real UIB emission has to rely on the detailed shape of the 8$\mu$m
peak and presence of the other UIB features.

Table~\ref{tab:sources_rk} summarizes the presence of the UIB
bands. About half of the targets exhibit a spectrum with at least the
6.2, 7.7 and 11.3 $\mu$m mid infrared bands clearly visible.  The
remaining targets do not exhibits all these bands, and two none of
them significantly. It is striking that almost all the observed ULIRGs
show a strong feature at or near 7.7 $\mu$m. IRAS F23529-2919 is an
exception. Here, the continuum might be so strong that it can dominate
the emission of the 7.7 $\mu$m emission band. Also, the shape of the
7.7$\mu$m feature is very different in two other galaxies from the
typical UIB band.  IRAS 22192-3211 and IRAS 00275-2859 have a band
which is broader than the one observed in a ``normal'' starburst
galaxy. The signal to noise for IRAS 23515-2917 is lower so it is more
difficult to draw a definitive conclusion. A strong continuum has been
detected in the two former targets, which supports the presence of an
AGN in both sources.  The change of the shape of the 7.7 $\mu$m
feature can be explained in two ways. The carriers of the UIB might be
processed in the AGN environment, changing the shape of the spectral
emission of these carriers. Such changes in the shape of the
7.7/8.6$\mu$m complex have in fact been observed in some ultra-compact
\ion{H}{2} regions \citep{Cesarsky96a}. Alternatively, a ``fake'' 7.7
$\mu$m feature could arise when one observes a high optical depth source
with self-absorbed silicate emission. There is no agreement in detail, 
however, between the broadish 8$\mu$m features in these
ULIRGs and the spectral shapes predicted by torus models
\citep{pier92}. A further problem of a `torus' identification of these
8$\mu$m features is that they are not observed in the spectra of
template Seyferts and QSOs
\citep{rigopoulou99,clavel00}.

Another source, IRAS F00183-7111, has a maximum near 8$\mu$m which we
do not ascribe to UIBs. Its spectrum is remarkably similar to some
deeply obscured objects (d'Hendecourt et al. 1996, NGC 7538 IRS1 : 
D.Cesarsky unpublished), showing a strong 5-8$\mu$m continuum and an
extremely deep silicate absorption.  Even an apparent ``6.5 $\mu$m
emission'' is indicated in this object which is very likely the result
of the absorption of water ice at 6 $\mu$m and an absorption feature
at 6.85$\mu$m which is of unclear origin \citep{schutte98}. The
spectrum of IRAS F00183-7111 is thus dominated by a heavily absorbed
(AGN) continuum with a possible weak 7.7$\mu$m emission, which might
be either due to a classical UIB or self-absorbed silicate
emission. In total, it appears to be the only unambiguous example in
our data for a type of source considered typical in the classical
Sanders et al. (1988) scenario: an extremely obscured AGN.

\section{Quantitative diagnostics of the Starburst and AGN activity in ULIRGs}

The use of the line-to-continuum ratio of the 7.7 $\mu$m band
(hereafter L/C) as proposed by Genzel et al. (1998) and Lutz et
al. (1998) to distinguish in the {\it mid-infrared} between
starburst-dominated and AGN-dominated ULIRGs is based on evidence that
a starburst galaxy presents strong UIB features while AGN presents
strong continuum with a low contrast emission features in the
nuclear part of the galaxy. This is suggested by ground based
observations \cite{moorwood86,roche91}, ISO spectroscopy
\cite{genzel98,rigopoulou99}, and ISO spectro-imaging
\cite{moorwood99,laurent99}.  This also implies that the UIB
luminosity of a galaxy scales with its star formation.  Recent studies
tend to reinforce this idea : in spiral galaxies, the global H$\alpha$
emission of the disk is correlated with the total MIR emission of UIB
\cite{vigroux99,roussel2000}.

When using these features as tracers of the sources of the total ULIRG
luminosity, a further assumption is that the mid-infrared actually
probes the active regions and not just a surface layer.  This is not
trivial since the active regions of ULIRGs are small
\cite{condon91,soifer2000} and dusty. Depending on dust configuration,
a major part of the luminosity could be hidden from mid-infrared view.
Genzel et al. (1998) have addressed this problem using the
starburst-type mid-infrared nebular spectra of ULIRGs. For ULIRGs, the
ratio of the inferred starburst ionizing luminosity to the bolometric
luminosity approaches that for template starbursts, indicating that
more than half of the luminosity has been traced by the mid-infrared
observations. Similar analyses can be made using the luminosity of the
PAH features.  Rigopoulou et al. (1999) and Genzel \& Cesarsky (2000)
find that observed PAH luminosities of ULIRGs are well above those of
lower luminosity sources, the ratio to bolometric luminosity being
only a factor 1.4-2 lower than in starbursts. This factor is plausibly
explained by the higher extinction in ULIRGs and perhaps the effect of
the more intense radiation field on the PAH emission. We have
tentatively repeated this comparison using the extinction correction
derived from our fits described below, which however give only fairly
uncertain extinction for individual noisy spectra. The mean
discrepancy narrows down to a factor 1.2. Thus, the mid-infrared
diagnostics are probing the major part of the luminosity.

For a narrow line and emission from a single region, both line and
continuum are obscured by the same amount.  The width of the PAH
feature and the strong wavelength dependence of extinction at the
onset of the silicate feature make the PAH L/C less ideal. In some
objects the absorption might be so strong that it can mimic a 7.7
band. There are, however, differences in shape and the L/C will
normally stay below the cutoff of 1 adopted in previous papers. IRAS
F00183-7111 is the only clear example in our sample. Another technical
point that has to be verified is the determination of the continuum
under the 7.7$\mu$m feature. In the L/C diagnostic, this continuum is
deduced from the linear interpolation of the continuum levels at 5.9
and 10.9$\mu$m, i.e. the L/C ratio is affected by the flux
uncertainties at these wavelengths. This is also a systematic source
of uncertainty since the 10.9$\mu$m point is still within the silicate
feature and does not probe a clean continuum.

We have explored alternative methods in order to verify the robustness
of the L/C ratio with respect to the systematic error induced in the
7.7$\mu$m continuum determination.  First, we constructed the ``narrow
band'' ratio between the flux density at 7.7 and 6$\mu$m. Strong UIB
emission will create a high ratio which is not reached by a pure
continuum unless it is unrealistically steep over this small
wavelength interval. This diagnostic should hence identify strong UIB
emission easily but will face problems discriminating weak UIBs from
continuum slope variations.  The strength of this diagnostic is to
avoid any assumption on the shape of the spectrum in the region
obscured by the silicate feature. Figure~\ref{LC_NB} shows the
correlation between the L/C and F(7.7)/F(6) diagnostics seen for
ULIRGs and template spectra observed by PHOT-S
\citep{rigopoulou99}, SWS \citep{sturm00}, and CAM-CVF (this
paper). The dispersion in this plot is explained by the variation in
continuum slope and the systematic error of the L/C discussed before.
The good correlation of the two diagnostics supports the use
of the L/C diagnostic to identify systems that are clearly
starburst-like and show strong UIBs. By definition of the F7.7/F6
diagnostic, Figure~\ref{LC_NB} cannot test the effect of continuum
placement uncertainty in the L/C method on measurements of faint
residual UIBs in AGN-like systems. The line
F(7.7$\mu$m)/F(6$\mu$m)=L/C+1 is also overplotted in
Figure~\ref{LC_NB} to indicate the asymptotic case where the continuum
is flat.

To investigate further how the different spectral components influence
the MIR spectrum of a galaxy, we created another diagnostic whose
basis is a more physical parameterization of the spectrum.

\subsection{Presentation of the model \label{pres_mod}}

Our aim is an empirical fitting procedure using the complete
mid-infrared low resolution spectra to derive the contribution of the
AGN and starburst components to the mid-infrared spectrum. In a second
step, results will be extrapolated to the bolometric luminosities.
The fit is based on template spectra of dusty starbursts and AGN, and 
applied to the whole set of ULIRG and template spectra
observed by PHOT-S \cite{rigopoulou99}, SWS \cite{sturm00}, and
CAM-CVF (this paper). Since absorption plays an important role for the
spectra of dust-rich ULIRGs, we incorporate the effects of absorption
that is higher than in the templates. In principle, explaining the
mid-infrared spectra of ULIRGs is a problem of three dimensional
radiative transfer requiring a consistent treatment of dust, ices, and
PAH features. Considerable insight can be gained, however, from a
simple superposition of a screen-obscured starburst and a
screen-obscured AGN. Arguments in favour of this approach are:
 (1) The mid-infrared diagnostic reaches most of the active region. 
     Unknown more obscured sources will be minor contributors.  
 (2) The use of an observed AGN template implicitly accounts for 
     warm dust close to the AGN. Screen extinction is a reasonable 
     first approximation for the effect of even more dust around the 
     AGN, since this dust will mostly just absorb in the mid-infrared 
     consider here, and reemit in the far-infrared.
 (3) An aggregate of star forming regions is in principle better 
     represented by a case where emission and absorption are well
     mixed. Mixed case extinctions are difficult to determine, 
     however, unless there are probes measured at high S/N over a 
     large range of optical depths, a requirement not met by our spectra. 
     The screen case adopted here will give similar results for 
     optical depths around one, fall short at higher optical depth, but 
     be conservative in the sense of not inferring excessive starburst 
     activity in uncertain cases.
      
\subsubsection{The Starburst component}

Genzel et al. (1998), Rigopoulou et al. (1999) and Sturm et
al. (2000) showed that in the range of the 6-9$\mu$m UIB bands there
is little variation in starburst MIR spectra observed by SWS and
PHOT-S. Starburst spectra present some variations at longer
wavelength, however. Vigroux {\it et al.} (1996), noted that in some
regions of NGC 4038/4039 the flux at 15 $\mu$m can rise dramatically and
dominate the mid infrared spectrum. It can be shown that the MIR
spectrum of starburst galaxies is the superposition of two components
: one from photodissociation regions (PDR) and one from \ion{H}{2}
regions \citep{tran98}. The PDR component is responsible for the UIB
emission and the \ion{H}{2} region for the hot dust continuum which
becomes important at 12 $\mu$m
\citep{Cesarsky96a,verstraete96,roelfsema96}. This hot dust continuum
is probably due to very small grains heated to temperature as high as
200K in some regions. This continuum is very faint at 7$\mu$m
\citep{Cesarsky96a,verstraete96,tran98}, and is not
really noticeable in the spectrum of the majority of starburst
galaxies where the PDR contribution ({\it i.e.} the UIB contribution)
dominates in this part of the spectrum \citep{sturm00}. For some dwarf
metal-deficient starburst galaxies like NGC 5253 and IIZw40
(Rigopoulou et al. 1999, S.Madden private communication), UIBs are
nearly absent.  In such a case a strong featureless continuum will not
be an AGN signature. Since ULIRG are dust rich and not known to be
metal-deficient this effect is likely not relevant, and we still
consider a strong ULIRG continuum with weak UIBs an AGN signature.

As the (rest)-wavelength range investigated in this
paper is between 5 and 12 $\mu$m, the contribution of the hot dust
continuum from \ion{H}{2} regions will not change drastically the
spectrum of the starburst contribution. But one {\it must} keep in
mind that strictly speaking a significant continuum at short
wavelengths (5-8 $\mu$m) is not necessarily a tracer of an AGN, but
could reveal a higher weight on \ion{H}{2} regions in the ISM with
respect to photodissociation regions. To break the degeneracy, one has
to look at the continuum at 15 $\mu$m since the continuum of an
\ion{H}{2} region is steeper than the continuum of an AGN
\citep{laurent99}.

In the present model we assume that the template spectrum of a
starburst does not change in a significant way between 5 and 12 $\mu$m
(see section 4.2 for the self consistency of this hypothesis). {\bf We
have used the M82 SWS spectrum as starburst template spectrum},
supposed to represent the standard line to continuum ratio of the UIB
seen in a dusty starburst.

\subsubsection{The continuum component}

Nuclear AGN spectra have low contrast UIB features and mainly
present a continuum emitted by hot dust. Obviously, when looking at
the total spectrum of a galaxy hosting an AGN, one naturally observes
UIB bands from circumnuclear star formation or from the disk of the
AGN host (see for instance
\citep{mirabel99,clavel00,moorwood99,alexander99}).  {\bf The AGN
spectrum itself can be described by an intrinsic continuum which can
be well fitted by a power law\footnote{The `power law' does not
mean non-thermal emission, and thus it is a parameterization of the dust
continuum.} of variable index} \citep{clavel00} subject to absorption
by dust especially in the region of the 9.7$\mu$m silicate absorption
feature.

As discussed in the previous subsection, a continuum seen in the
mid-infrared may also be due to \ion{H}{2} regions. In the wavelength
range considered (5-12 $\mu$m), there is a degeneracy between the AGN
continuum and a continuum coming from an \ion{H}{2} region, although
an \ion{H}{2} continuum is usually steeper and fainter. Our fit
invokes a single `continuum component' including both possibilities
of an AGN continuum and an unusually strong \ion{H}{2} continuum. We
will come back to the degeneracy between these two types of continua
when discussing fit results.

\subsubsection{Dust obscuration}

Because ULIRGs are gas and dust rich galaxies \citep{solomon97}, the
extinction is expected to play a crucial role on shaping the spectrum
of these galaxies, even in the mid-infrared (see for instance Arp220,
\cite{sturm96}).  It is important to test how dust obscuration can
affect both UIB dominated and featureless spectra
\citep{rigopoulou99}. One can argue that mid-infrared ULIRG spectra
could be completely dominated by absorption, and that the 7.7$\mu$m
feature, which we identify as an UIB, could result from a deep
absorption of a flat continuum \citep{dudley97}. This could be the
case in the strongly absorbed galaxy F00183-7111. We want to test more
generally whether it is possible to produce an emission feature around
8 $\mu$m by a power law continuum affected by an absorption law or
whether a simple UIB spectrum affected by absorption is preferred.

We used two absorption laws, the `classical' Draine and Lee (1984)
absorption law and one composite absorption law derived from the
Galactic Center-SgrA* spectrum (hereafter GC or GC-SgrA*).  The GC
extinction law has a higher extinction level between 3 and 7$\mu$m
\citep{lutz96,lutz99a}. Since this law is derived from hydrogen
recombination lines and does not sample the silicate feature well, we
have supplemented it with points deduced from the GC continuum
spectrum itself.  They have been computed assuming that the intrinsic
GC spectrum is a power law spectrum absorbed by a dust screen. The
extinction law and the exponent of the power law have been adjusted in
order to fit the extinction deduced from the hydrogen lines and the GC
spectrum itself. This procedure is adequate since we cover only a
limited wavelength range, and since the dust in front of the Galactic
center is close to the ideal screen case.  This absorption law has two
main peculiarities: a flatter absorption between 3 and 7$\mu$m, and a
narrower silicate feature than what the Draine and Lee curve predicts
(see Figure ~\ref{GC_ext}).

These two absorption laws have been used to represent variations in
dust properties as possible in extreme environments such as close to
an AGN. Comparison of the M\,82 and NGC\,1068 SWS spectra
\citep{sturm00} suggests such variations at least for absorption {\em
features}. Because of the possibility of such variations, absolute
A$_V$ values derived from the model are somewhat uncertain. The main
emphasis is to test whether the shape of the ULIRG spectra can be
reproduced well.

\subsubsection{The different models}

The two spectral components and obscuration have been used to
construct four types of models in two classes:

\begin{itemize}
\item 
The first class is based on the assumption that the starburst
component and the continuum component are obscured by the same dust
screen. Models of this class vary in extinction properties: the dust
extinction law is either a ``pure'' Draine and Lee law, or a ``pure''
GC extinction law or a mixture of the two.
\item
The second class allows different obscuration of the two components
and is clearly more realistic. The two different contributions
(continuum and starburst) are not necessarily located in the same
physical region and therefore do not suffer the same extinction.
\end{itemize}

Table~\ref{tab:mod_descr} provides the formal description of the
different models used. Models 1a, 1b and 2 refer to the first class,
while models 3a-d and 4 refer to the second class.  In two
cases (2 and 4), the extinction law is a linear combination of
the GC extinction law and the Draine and Lee one.  This combination is
not constrained by observations but included in our modeling as a
comparison to other models with different degrees of freedom.

\subsection{Testing the validity of the model }

To validate the model, the fit quality is quantified first.
Second, the results from test fits to template starburst and AGN spectra
are investigated, and compared to
observations of ``mixed'' sources where the AGN and starburst relative
contributions are well known from spatially resolved data.

\subsubsection{Quality of model fits}

Figure~\ref{disp_fit} shows some examples of fits to ULIRG spectra
using model 3a. The figure includes AGN and starburst-like spectra of
both high and low S/N. Fits using other models tend to be slightly
worse but still acceptable. The first conclusion is that the
models presented in Table~\ref{tab:mod_descr} can give a reasonable
fit for all ULIRG spectra. This implies that the assumptions of these
models ({\it e.g.} a MIR spectrum of a ULIRG is a superposition of a
continuum component and a starburst component extincted by a dust-rich
ISM) are not in obvious contradiction with observations.

To better quantify the quality of the various models we adopt the
indicator $\frac{\chi^2}{N_{\rm free}} = \frac {{\sum(F_{\lambda}^{\rm obs}-
F_{\lambda}^{\rm mod})^2}}{{\sum(\sigma^{\rm obs})^2}} \frac{1}{N_{\rm free}}$
where $F_{\lambda}^{\rm obs}$ is the observed flux density,
$F_{\lambda}^{\rm mod}$ is the model flux density, ${\sigma}^{\rm obs}$ is the
rms of the observed flux density.  The summation is done over 
{\em all} ULIRG spectra, and $N_{\rm free}$ is the sum of degrees of
freedom of modeled spectra. With this indicator the quality of the 
different models can be compared: the closer to 1., the better.
Table~\ref{tab:comp_stddev} presents the mean
 $\frac{\chi^2}{N_{\rm free}}$ both for our complete ULIRG sample, and for
 a high quality subset with observed S/N greater than 3.

\subsubsection{Comparison with template sources.}

We have obtained model fits for various starburst, AGN, and mixed
template sources. A first result can be obtained from a comparison of
the two classes of models for the mixed templates Circinus and
Centaurus A. Figure~\ref{comp_visu} shows the obvious superiority of
models where the extinction towards the AGN is different from that
towards the starburst.  Fully equivalent good fits are achieved for
models 3a-3c.

Figures ~\ref{verif_Stb} and ~\ref{verif_Agn} show the observed
spectrum and the fitted spectrum computed with model 3a, for several
starburst and AGN templates. The Galactic center has been grouped with
the AGNs since the very center shown here is also a continum source
devoid of UIBs \citep{lutz96}. The absence of these features in the
GC-SgrA* spectrum is due to the presence of an \ion{H}{2}-region with
an intense radiation field that probably destroys the UIB carriers.
We overlay the contributions of the starburst component and the
contributions of the continuum component.  As expected, the starburst
component dominates totally the modelled spectrum for the starburst
galaxies. Similarly, a quasi-pure continuum contribution is needed to
fit the AGN spectra, with extinction being negligible in many cases
but significant in others (NGC 5506). The low metallicity galaxy NGC
5253 (not shown) is well fitted by a strong continuum contribution and
weak UIBs. The fits to the starburst templates (Fig.~\ref{verif_Stb})
show at their long wavelength end a varying contribution of a faint
steeply rising continuum. Most likely, this is due to variations in
the strength of the 15$\mu$m VSG (Very Small Grains) continuum, a
conclusion also reached by Sturm et al. (2000) on the basis of the SWS
spectra of M82 and NGC 253.

A further test is to apply the model to objects which are well
known to be composite, containing both starburst and an AGN. In
Figure~\ref{verif_mix} spectra of four such objects are presented,
showing plausible decompositions into the starburst and continuum
components. A stringent test can be obtained for Centaurus A and
Circinus which have both been observed by ISOCAM/CVF at high spatial
resolution (Mirabel et al. 1999; Moorwood et al. 1999 and in preparation). These
observations allow us to separate the contribution of the nucleus and
the surrounding star-forming regions, and thus a direct test of our
model decomposition.  The spectrum of the Centaurus A nucleus,
extracted from the ISOCAM-CVF observation, is overplotted in
Figure~\ref{verif_mix}. The modelled continuum contribution is indeed
very similar to the observed spectrum of the nucleus, confirming the
validity of our approach. We performed the same comparison for the
Circinus galaxy (A.F.M. Moorwood, private communication) and find the
same good agreement between the observed nuclear continuum and the
modelled continuum contribution.  At least for these two sources,
the separation into two contributions (continuum and starburst) is reliable.

\subsection{Results of the modelling of starburst and continuum 
contributions \label{res_mod}}

Table~\ref{tab:comp_stddev} shows that the two classes of models do
not reproduce the ULIRG spectra equally well. It is obvious that the
models of the second class, where extinctions of continuum and
starburst are different, are in better agreement with the
observations. This better accuracy is not explained trivially by the
increasing number of free parameters. Model 2 which has the same
number of free parameters as the `good' models 3a-c is clearly less
successful.  That means, more flexibility in the shape of the
extinction does not improve the fit significantly, while invoking
different extinction towards the starburst and AGN does.  This lends
support to the plausible idea that two different extinctions are
applicable to the physically distinct starburst and AGN parts of a
ULIRG. Comparing the continuum and starburst extinction derived by our
fit procedure, we find the continuum extinction to be slightly smaller
on average.  This is plausible if the obscuring dust is not
distributed with spherical symmetry. We note, however, that this
result should be considered tentative given the considerable
uncertainty of individual A$_v$ values.

Figure~\ref{LC_RATAS} compares the L/C ratio used in previous papers
with the ratio of the integrated 5-10$\mu$m fluxes contained in the
continuum contribution and in the starburst contribution as deduced
from the model 3a.  In this plot we also add the starburst and AGN
comparison samples as described by Rigopoulou et al. (1999).  The two
indicators are well (anti)correlated. This correlation shows {\it a
posteriori} the consistency of the two indicators. 
Table~\ref{tab:ulirg_ratas_lc} displays the
individual value of L/C and the ratio of the integrated 5-10$\mu$m
fluxes contained in the continuum contribution and in the starburst
contribution.

Genzel et al. (1998) and subsequent papers adopted a L/C threshold of
1 to discriminate between starburst and AGN dominated systems. At this
threshold, the flux ratio of the continuum and starburst components,
each integrated over the 5-10$\mu$m range, will be larger than one,
simply because the UIBs are relatively narrow features. This is
reflected in Figure~\ref{LC_RATAS}. The purpose of the threshold,
however, is to identify the component dominating the bolometric
luminosity and not just the 5-10$\mu$m flux. The mid-infrared
contributions have to be extrapolated to the bolometric
contributions. Using the spectral energy distributions of M82 and of
the central region of NGC\,1068 as starburst and AGN templates (Sturm
et al. 2000 for M82, including IRAS points and correcting for aperture
effects), we find the ratio $L_{5-10\mu\/m}/L_{\rm IR}$ about 2.5 times
greater in the case of the AGN than for starbursts. Very similar
values are obtained for other AGN templates (NGC 4151, 3C273),
consistent with the notion of stronger mid-IR continua in Seyferts
than in starbursts \citep{degrijp85}. Orientation related variations
in mid-infrared Seyfert continua \citep{clavel00} caution that there
will be some scatter in these bolometric corrections for individual
AGN.  In the remainder of this paper, we will apply the factor of 2.5
wherever total infrared (i.e. bolometric) quantities are
discussed. Taking into account the correction factor 2.5,
Figure~\ref{LC_RATAS} confirms the L/C threshold of 1 as a sensible
one for differentiating between sources dominated by the starburst
component and those dominated by the AGN component.

\section{Discussion}

\subsection{Mid-infrared spectra and power sources of the highest 
luminosity ULIRGs}

Determination of the power source of the highest luminosity ULIRGs is
a prime goal of our project, extending the ISOPHOT-S sample which is
dominated by lower luminosity sources \citep{lutz98,rigopoulou99}.
Figure~\ref{LC_comp} displays the L/C ratio for the CAM-CVF
ULIRGs together with data from Lutz et al. (1998) and the two
hyperluminous sources observed by ISO \citep{taniguchi97,aussel98}.
The UIB L/C is used to classify sources as starburst-like if
L/C~$>$1 or AGN-like if L/C~$<$ 1. From our enlarged database, we
confirm that most ULIRGs are starburst dominated in the mid-infrared
but that there is a trend towards AGN-like ULIRGs at higher
luminosities. In our sample which contains 17 sources above
$L_{\rm IR}=10^{12.4} L_\odot$, the highest luminosity source classified
as starburst is at $10^{12.65} L_\odot$ (IRAS 03835-6432). Such a
luminosity limit for the `most luminous starburst' is in agreement
with the timescale argument given by Heckman (1994). Adopting the
bolometric luminosities of our sources and sizes like those derived by
Soifer et al. (2000) for the mid-IR emitting region of ULIRGs, the
surface brightness will exceed the limit for starbursts suggested by
Meurer et al. (1997).  Individual star forming region are known to exceed
this limit, however, and the huge gas concentration in the centers
ULIRGs provides sufficient fuel.  Additionally, Meurer et al. (1997)
derived their limit using, for similar sources, H$\alpha$-based sizes
which are larger than the mid-IR ones found by
Soifer et al. (2000), thus staying below the actual surface brightness.
Even in our sample enriched in luminous ULIRGs, statistics at the
highest luminosities is limited, thus maintaining the possibility for
higher luminosity starbursts. Given the modest sample size, it is
difficult to assign an accurate value for the luminosity at which half
of the sources are starburst and half AGN dominated. From
Figure~\ref{LC_comp}, this occurs most likely at luminosities
$10^{12.4} L_\odot$ to $10^{12.5} L_\odot$.

The trend from starburst dominance for most `normal' ULIRGs to AGN
dominance at the highest luminosities now consistently emerges from a
variety of indicators: Mid-infared continuum \citep{sanders88a},
mid-infrared spectroscopy (Lutz et al. 1998 and this paper), and
optical spectroscopy \citep{veilleux99}. As noted by \citep{lutz99},
the good agreement between optical and mid-infrared suggests that AGN
in ULIRGs do not remain fully embedded for long but manage to break
the obscuring screen at least in certain directions.

The highest luminosity `starburst-like' systems appear well above our
cutoff L/C of 1, but below the L/C$\sim 3$ typical for starbursts and
lower luminosity ULIRGs. This can be interpreted in different ways.
Either they contain a noticeable though not dominant AGN component, or
the increased continuum contribution is due to an increase in the
\ion{H}{2}-region related continuum, as might be reasonable for 
sources with very high star formation rates in a small region.

\subsection{The nature of the mid-infrared continuum}

The simple L/C method will not be able to break the degeneracy for
the origin of ULIRG mid-infrared continua. Also, the parameterization
chosen for our fits (Sect. \ref{pres_mod}) describes the spectrum of a
ULIRG in terms of just two components : a UIB-dominated starburst
component and a continuum component which is either due to an AGN or
extra \ion{H}{2} region contributions. If the relative weight of
`PDR-like' and `\ion{H}{2} region-like' components varies for
different starburst conditions, as seen e.g. for different regions of
the Antennae galaxies \citep{vigroux96}, assigning the continuum
entirely to an AGN may be misleading. The nature of the continuum has
to be investigated in more detail to quantify the fraction of energy
radiated by the AGN (AGN continuum) and by star formation activity
(UIB-dominated component plus \ion{H}{2} region continuum).

It is possible to break the continuum degeneracy by using the slope of
the continuum (Laurent {\it et al.}  1999). The ratio of the fluxes at
6$\mu$m and 15 $\mu$m is different for an \ion{H}{2} region and a
continuum emitted by the dust (torus) surrounding an AGN. Dust in the
inner parts of \ion{H}{2} regions emits a mid-infrared spectrum that
is steep over our wavelength range.  This can be seen both in Galactic
\ion{H}{2} regions \citep{verstraete96,cesarsky96} and in extragalactic
regions of particularly concentrated star formation, e.g. the star
forming knot A in the Antennae galaxies \citep{vigroux96,mirabel98}.
In contrast, the continuum emitted by hot AGN dust is much flatter in
the mid-infrared, as shown by the examples of Cen A \citep{mirabel99}
and NGC1068 \citep{lutz00}.

Because of the presence of the UIB emission and silicate absorption
features, the steepness of the continuum can be determined reliably
only with sufficient wavelength coverage, ideally from 6$\mu$m or less
to 15 $\mu$m (rest wavelength).  This condition is not met by the
ISOPHOT-S spectra which form most of the ISO database of low
resolution ULIRG spectra but cut at less than 11$\mu$m rest wavelength
for most ULIRGs. It is partially met for the ISOCAM-CVF spectra
presented in this paper which typically extend to rest wavelength
about 12$\mu$m.  It is instructive to discuss the case of Arp 220 for
which a full CVF spectrum is available \citep{charmandaris97}. Arp 220
shows 6.2 and 7.7$\mu$m UIBs on no or weak continuum, whereas the
continuum in the 11-15$\mu$m region is clearly stronger than in a
standard M82-type starburst spectrum. This suggests a strong
but steep \ion{H}{2} continuum as in the star forming knot A of the
Antennae, subject to even higher extinction (see
Figure~\ref{arp220_knotA_comp}).

The case is less clear for our spectra which do not extend to
15$\mu$m. Some spectra suggest a strong but steep continuum at the
long wavelength end of a UIB dominated spectrum (e.g. IRAS 03521+0028)
while others are fit by a UIB dominated spectrum plus a flat continuum
(e.g. IRAS 18030+0705). At the limited wavelength coverage, the
uncertainties are considerable, also taking into account that fits are
affected by some 7-8$\mu$m features not showing the typical PAH shape
(Sect. \ref{results}).

\subsection{Model results: Contributions of starburst and AGN activity}

Our model can be used to better quantify the contribution of
starburst and AGN activity independent from the L/C criterion used
earlier.  We measured the mid infrared fluxes emitted by starburst and
continuum contribution (not dereddened) integrated over the wavelength range
5-10$\mu$m (similar to the filter LW2 of ISOCAM). When discussing
total infrared (bolometric) contributions instead of 5-10$\mu$m ones,
the different AGN and starburst SEDs are considered by
increasing the starburst contribution by the correction factor 2.5
derived in Sect. \ref{res_mod}. Since we were not able to reliably
break the degeneracy of AGN and \ion{H}{2} region continuum for our
wavelength coverage and signal-to-noise ratio, we explored two extreme
cases for the origin of the continuum,  reality most likely being in
between.

The first scenario assumes that the continuum is coming entirely
from an AGN. The examples of \ion{H}{2} region continua discussed
above, however, clearly argue for {\em some} star formation
contribution to the continuum. It is unlikely that an almost
pure continuum spectrum is produced in this way. The only
star-formation powered galaxies approaching this case are low
metallicity dwarfs like NGC 5253, an unlikely match to the ULIRG
situation. We hence have adopted the following second extreme
scenario: All faint continua, defined as emitting less than 50\% of
the total 5-10$\mu$m flux, are assumed to be due to star formation,
while all stronger ones are due to an AGN.

Figure~\ref{Lir_mirfrac} shows the median ratio of the energy emitted
by the starburst component to the total energy emitted
($F_{5-10{\mu}m}({\rm Starburst})$/$F_{5-10{\mu}m}({\rm Total})$) of
ULIRGs presented in Table~\ref{tab:ulirg_ratas_lc} and two HYLIRGs
\citep{taniguchi97,aussel98}, versus the total infrared luminosity.
A trend towards AGN dominance at high luminosity is again clearly
seen.  The cutoff between starburst-dominated galaxies and
AGN-dominated galaxies lies at $L \approx 10^{12.4-12.5}L_{\sun}$.
The significance of this trend can be tested by dividing the
luminosity range into the bins $L_{\rm FIR}=10^{11.8-12.4} L_\odot$
and $L_{\rm FIR}=10^{12.4-13.0} L_\odot$ and using respectively the
Kolmogorov-Smirnov's test and the Student's t-test to test if the two
samples follow the same distribution or have the mean value. The
probability for such hypotheses are $1.0 \times 10^{-4}$(resp. $1.3
\times 10^{-4}$), demonstrating that the trend is statistically very
significant. We also perform the two statistical test while removing
the two most luminous galaxies ($L_{\rm FIR} > 10^{13} L_\odot$).
Probabilities are then $9.0 \times 10^{-4}$ and $5.0 \times 10^{-4}$.

To quantify the AGN contribution at
different luminosities, we identified either all or none of the
continuum of each source as AGN according to the two scenarios
described previously. We then determined the ratio
$F_{5-10{\mu}m}({\rm Starformation})$/F$_{5-10{\mu}m}({\rm Total})$, with
$F_{5-10{\mu}m}({\rm starformation})$ being the sum of UIB-dominated
starburst flux and that part of the continuum flux that is ascribed to
star formation under the given scenario.  For each scenario, we
averaged the results for the ULIRGs in each of the two luminosity
bins. Table~\ref{tab:RATAS_Lir} displays the resulting average values
along with their uncertainties. The average contribution of star
formation to the 5-10$\mu$m flux is 65 to 84\% for the low luminosity
bin and 24 to 33\% for the high luminosity bin, the ranges
representing the two extreme scenarios. Extrapolating from
mid-infrared fluxes to total far-infrared ones, the average
contributions of star formation are 82 to 94\% for the low luminosity
bin and 44 to 55\% for the high luminosity bin. These results from
quantitative modeling considerably reinforce the conclusions from the
earlier work based on the L/C ratio: Low luminosity ULIRGs are
predominantly starburst powered, while AGN powering dominates above
$\sim 10^{12.4}$ to $10^{12.5} L_{\sun}$.

\section {Conclusions}

We have presented ISOCAM-CVF low resolution mid-infrared spectroscopy
of 16 ultraluminous infrared galaxies obtained within a
program targeted at investigating the most luminous
ULIRGs. The data are analyzed along with the complete ISO database of
low resolution spectra of ULIRGs/HYLIRGs, totalling 76 sources. We use
the presence of the mid-infrared `UIB' aromatic bands as a diagnostic
of starburst activity, finding starburst-dominated systems up to a
luminosity of $ L_{\rm IR}=10^{12.65}L\odot$.  Other spectra show a strong
AGN continuum with weaker features of uncertain origin. We have found
one highly obscured AGN.

The mid-infrared spectra can be modeled in terms of a superposition of
a UIB-dominated starburst spectrum and a continuum, both potentially
strongly obscured. This continuum contains both AGN and additional
\ion{H}{2} region contributions. The fits prefer a two zone model in which
the extinction to starburst and continuum regions may differ.  Results
from these fits agree well with previous simpler diagnostics based on
the line-to-continuum of the UIB features.

Low luminosity ULIRGs are mostly starburst dominated, but the AGN
fraction increases with luminosity and most ULIRGs above $\sim
10^{12.4}$ to $10^{12.5} L_{\sun}$ appear AGN like.  We have
separated ULIRGs into two luminosity bins of $L_{\rm FIR}=10^{11.8-12.4}
L_\odot$ and $L_{\rm FIR}=10^{12.4-13.0} L_\odot$. We find that the
average contributions of star formation to the infrared luminosity are
82 to 94\% for the low luminosity bin and 44 to 55\% for the high
luminosity bin.

\acknowledgments

We are grateful to Michael Rowan-Robinson for allowing us access to
the QDOT ULIRGs prior to publication.  The work of QDT is supported by
Verbundforschung (50 OR 9913 7).  DCH acknowledges support from NASA
grant NAG5-3359 to US ISO Observers.  DR acknowledges support of the
EC TMR Network "Infrared Surveys" (under contract number
ERBFMRX-CT96-0068).  YT is supported by the Ministry of Education,
Science, Sports and Culture of Japan under grants 10044052, and
10304013.  SWS and the ISO Spectrometer Data Center at MPE are
supported by DLR (DARA) under grants 50 QI 8610 8 and 50 QI 9402 3.

\appendix
\section{Reduction scheme for CVF observations of faint point sources 
\label{PSF_red} }

Aiming for a good CAM-CVF spectrum including a reliable continuum, a main
source of uncertainty is due to the Point Spread Function 
(PSF). Because the width of the PSF
changes with the observed wavelength, it is crucial to determine the
observed total flux for each wavelength separately. This is easily
done for bright point sources because it is easy to sum the observed
flux over a large box (typically 5 by 5 pixels) or to fit the PSF
profile to the observed image.  Since our sources are point-like at
CAM spatial resolution, it is not necessary to consider the complex
situation for extended sources.  Even for point sources, however, the
PSF correction is difficult if they are faint. Then, only a few pixels
around the brightest one have a significant S/N (hereafter the
brightest pixel will be called ``central pixel'').  In such a case,
the S/N of the corrected spectrum obtained by summing over a box or
fitting the PSF profile is usually very low.

A better S/N can be obtained just using the signal of the central
pixel and applying a PSF correction factor that varies slowly with
wavelength.  The main difficulty in determining this factor is to
obtain the position of the source within the central pixel. This could
be done by fitting a model of the PSF to the observed images at each
wavelength. But, because the S/N is quite low for the pixels around
the central pixel, direct application of this method gives uncertain
positions and a final spectrum where the S/N is low, comparable to the
S/N obtained by summing over a box.  In order to improve the S/N we
adopted the following method :

In CIA, theoretical PSFs are available at every half micron of the
ISOCAM-CVF wavelength range, residual variations over such ranges are
small. We divided the observed wavelength range
into sub-ranges of one micron centered around the wavelength of a
theoretical PSF, and summed images corresponding to wavelengths in
this range of one micron around the given wavelength (i.e. about 10
images). We then fitted a PSF profile to the resulting image
(hereafter the summed image). This method improves the S/N, and the
deduced position of the source in the central pixel is known with a
better confidence level. We used this position and the PSF to derive
the correction factor from flux in the 'central' pixel to total flux,
which was applied for each image in the considered wavelength range.

Although this method gives a better position of the source, it is still
ncessary to compute the systematic error introduced by this correction.
The formal way is to compute the S/N for each pixel of the summed
image, and then to use an inverse Monte-Carlo method to compute the
S/N of the spectrum. Another equivalent way which avoids a Monte-Carlo
simulation is to consider the images used to obtain the summed image
{\it as guesses of a Monte-Carlo simulation }. This is equivalent to a
simulation with ten guesses because the S/N of the summed image is
deduced from these images.  We then used the positions of the source
obtained for all these images to correct the flux of the summed image.
>From the scatter of these measurements, it is then possible to deduce
the systematic error introduced by the fitting method of the PSF
profile.

\begin{table}
\caption{Basic data of ULIRGs observed with ISOCAM-CVF}
\begin{tabular}{lrrcrrrrr}
\tableline
\tableline
IRAS Name\tablenotemark{a}&RA(2000)
&DEC(2000)&z &$S_{12}$&$S_{25}$&$S_{60}$&$S_{100}$&Log($L_{\rm IR})$\\ & &
& &Jy &Jy &Jy &Jy &$L_{\sun}$\\
\tableline
F00183-7111& 0:20:35.3&-70:55:22 &0.327& $<$0.06&    0.13&    1.20&    1.19 
&12.77\\
 00188-0856& 0:21:26.6& -8:39:22 &0.129& $<$0.12&    0.37&    2.59&    3.40 
&12.31\\
 00275-2859& 0:30:04.4&-28:42:23 &0.279& $<$0.08&    0.17&    0.69&    0.73 
&12.46\\
 00406-3127& 0:43:03.2&-31:10:50 &0.342& $<$0.06& $<$0.09&    0.72&    0.99 
&12.64\\
 02113-2937& 2:13:33.0&-29:23:35 &0.194& $<$0.08& $<$0.12&    0.94&    1.88 
&12.29\\
F02115+0226& 2:14:10.3&  2:40:00 &0.400& $<$0.11& $<$0.16&    0.32&    0.64 
&12.48\\
F02455-2220& 2:47:51.3&-22:07:38 &0.296& $<$0.08& $<$0.10&    0.82&    1.27 
&12.57\\
 03000-2729& 3:02:11.5&-27:07:24 &0.221& $<$0.07& $<$0.11&    0.92&    2.04 
&12.41\\
 03538-6432& 3:54:25.3&-64:23:39 &0.310& $<$0.06&    0.06&    0.99&    1.30 
&12.65\\
 03521+0028& 3:54:42.4&  0:37: 8&0.152& $<$0.25&    0.23&    2.64&    3.83 
&12.46\\
 04384-4848& 4:39:50.8&-48:43:11&0.213& $<$0.04&    0.07&    0.99&    1.34 
&12.32\\
 17463+5806&17:47:04.6& 58:05:24&0.309& $<$0.04& $<$0.04&    0.65&    0.95 
&12.48\\
 18030+0705&18:05:32.5&  7:06: 9&0.146& $<$0.25& $<$0.25&    0.84&    4.40 
&12.18\\
 22192-3211&22:22:09.6&-31:56:34&0.231& $<$0.12&    0.19&    0.89&    1.42 
&12.43\\
 23515-2917&23:54:06.7&-29:00:58&0.336& $<$0.08& $<$0.14&    0.65&    1.06 
&12.60\\
F23529-2919&23:55:33.3&-21:03: 7&0.430& $<$0.08& $<$0.16&    0.33&    0.63 
&12.55\\ 
\tableline
\end{tabular}
\tablenotetext{~}{Positions have been computed from the ISOCAM
observations, with an accuracy of $\sim$5\arcsec.}
\tablenotetext{a}{A name starting with an 'F' refers to a FSC object
name. The others names are taken from the PSC.  
 }
\label{tab:sources}
\end{table}

\begin{table}
\caption{Detection of UIB features in the ULIRG spectra}
\begin{tabular}{lccccl}
\tableline
\tableline
Iras Name & 6.2 $\mu m$ & 7.7 $\mu m$ & 11.3 $\mu m$ & L/C & Note \\
\tableline
F00183-7111& n  & n  & n  & 0.88 &              \\
 00188-0856& n  & y  & y  & 1.53 &              \\
 00275-2859& n  & y  & n  & 0.58 &              \\
 00406-3127& -  & -  & -  & -    & Not acquired  \\
 02113-2937& n  & y  & y  & 2.53 &              \\
F02115+0226& n  & n  & n  & -    & No detection \\
F02455-2220& n & y  & n & -    & (see Sect. 2.2) \\
 03000-2729& n  & y  & y  & 1.74 &              \\
 03538-6432& y  & y  & n  & 1.70 &              \\
 03521+0028& n  & y  & y  & 3.23 &              \\
 04384-4848& y  & y  & n  & 2.08 &              \\
 17463+5806& n  & y  & n  & 1.85 &              \\
 18030+0705& y  & y  & y  & 1.88 & Large position offset\\
 22192-3211& n  & y  & n  & 0.85 &              \\
 23515-2917& n  & y  & n  & 1.36 &              \\
F23529-2919& n  & n  & n  & 0.38 &              \\
\tableline
\end{tabular}
\tablenotetext{~}{
y : feature is significantly present (detected at 
3$\sigma$ expect for F02455-2220)}
\tablenotetext{~}{
n : feature is doubtful or undetected} 
\tablenotetext{~}{
- : undefined}
\label{tab:sources_rk}
\end{table}

\begin{table}
\caption{Description of models used to fit ULIRG spectra and 
 quality of spectral fits}
\begin{tabular}{lcccc}
\tableline
\tableline
Model      & Type of model & Number of &$\frac{\chi^2}{N_{\rm free}}$&
$\frac{\chi^2}{N_{\rm free}}$\\
   & &parameters&  All ULIRGs& High S/N Ulirgs\\
\tableline
\tableline
1 a    & $ (\alpha_1  F_{\rm Stb} + 
\alpha_2  F_{\rm AGN} ) exp(-  \beta  A_{\rm DL})       $ &
4 & 2.48       & 4.79           \\
1 b    &  $(\alpha_1  F_{\rm Stb} + 
\alpha_2  F_{\rm AGN} ) exp(-  \beta  A_{\rm GC})       $  &
4 & 2.24       & 4.30           \\
\tableline
2 & $ (\alpha_1 F_{\rm Stb} + \alpha_2 F_{\rm AGN} ) 
exp(- \beta_1 A_{\rm DL} - \beta_2 A_{\rm GC}) $ &
5 & 2.23 & 4.31 \\
\tableline 
\tableline
3 a & $ \alpha_1 F_{\rm Stb} exp(- \beta_1 A_{\rm DL}) +
\alpha_2  F_{\rm AGN} exp(-  \beta_2  A_{\rm DL})  $ &
5& 1.70 & 2.81 \\
3 b    &  $ \alpha_1  F_{\rm Stb} exp(-  \beta_1  A_{\rm DL})    + 
\alpha_2  F_{\rm AGN} exp(-  \beta_2  A_{\rm GC})  $ &
5& 1.54       & 2.52            \\
3 c    & $ \alpha_1  F_{\rm Stb} exp(-  \beta_1  A_{\rm GC})    + 
\alpha_2  F_{\rm AGN} exp(-  \beta_2  A_{\rm DL})  $ &
5& 1.64       & 2.79            \\
3 d    & $ \alpha_1  F_{\rm Stb} exp(-  \beta_1  A_{GC})    + 
\alpha_2  F_{\rm AGN} exp(-  \beta_2  A_{\rm GC})  $ &
5& 1.72       & 3.05            \\
\tableline
4      & $ \alpha_1  F_{\rm Stb} exp(-  \beta_{1,1}  A_{\rm DL} - \beta_{1,2} 
A_{\rm GC})     $ & 
7& 1.42       & 2.21            \\
& $+ \alpha_2  F_{\rm AGN} exp(-  \beta_{2,1}  A_{\rm DL} - \beta_{2,2} A_{\rm GC} ) $ & & \\
\tableline
\tableline
\end{tabular}
\tablenotetext{~}{$F_{\rm Stb}$ represents the starburst template 
spectrum (M82), $F_{\rm AGN}$ the powerlaw continuum.  $A_{\rm DL}$ and
$A_{\rm GC}$ are the extinction laws of \cite{draine84} and of the
Galactic center, respectively. }
\label{tab:mod_descr} 
\tablenotetext{~} {Comparison of fit quality (see text) for the various 
models for all ULIRGs and for a high quality subset.}
\label{tab:comp_stddev}
\end{table}

\newpage
\begin{deluxetable}{lcc|lcc}
\tablecaption{Model results for ISOCAM-CVF and ISOPHOT-S 
targets\label{tab:ulirg_ratas_lc}}
\tablecolumns{6}
\tablehead{
\multicolumn{3}{c}{} &  
\multicolumn{3}{c}{} \\
\colhead{IRAS Name} & \colhead{Cont/Starb}\tablenotemark{a} & 
                    \colhead{L/C}\tablenotemark{b}    & 
\colhead{IRAS Name} & \colhead{Cont/Starb}   & \colhead{L/C}}
\startdata

 00153+5454 & 0.0 & 3.34 & 17028+5817 & 0.47 & 3.58 \\
F00183-7111 & 5.31 & 0.88 &  17068+4027 & 0.0 & 1.40 \\
 00188-0856 & 1.48 & 1.53 & 17179+5444 & 1.59 & 1.38 \\
 00199-7426 & 6.26 & 1.08 & 17208-0014 & 0.07 & 5.54 \\
 00275-2859 & 100. & 0.57 & 17463+5806 & 2.38 & 1.85 \\
 00397-1312 & 100. & 0.22 & 18030+0705 & 0.60 & 1.88 \\
01003-2238 & 100. & 1.44 & 18443+7433 & 6.09 & 1.70 \\
01166-0844 & 0.00 & 2.16 & 18470+3233 & 100. & 0.68 \\
01199-2307 & 0.00 & 0.97 & 18531-4616 & 1.59 & 2.59 \\
01298-0744 & 1.79 & 1.50 & 19254-7245 & 3.47 & 0.89 \\
01355-1814 & 0.00 & 0.93 & 19420+4556 & 0.0 & 3.69 \\
01388-4618 & 0.00 & 3.93 & 19458+0944 & 0.0 & 2.44 \\
01494-1845 & 0.00 & 1.95 & 20049-7210 & 0.0 & 3.77 \\
01569-2939 & 100. & 1.14 & 20100-4156 & 0.0 & 1.92 \\
02113-2937 & 0.64 & 2.53 & 20446-6218 & 0.0 & 0.94 \\
02364-4751 & 0.66 & 3.48 & 20551-4250 & 0.42 & 2.33 \\
02411+0354 & 0.18 & 3.17 & 21396+3623 & 0.0 & 2.31 \\
03000-2719 & 0.88 & 1.73 & 22055+3024 & 1.12 & 0.96 \\
03158+4227 & 0.53 & 1.72 & 22192-3211 & 100. & 0.84 \\
03521+0028 & 0.21 & 3.23 & 22491-1808 & 1.04 & 2.85 \\
03538-6432 & 1.45 & 1.70 & 23060+0505 & 100. & 0.08 \\
04063-3236 & 100. & 2.55 & 23128-5919 & 0.96 & 2.86 \\
04103-2838 & 100. & 1.37 & 23129+2548 & 0.0 & 3.55 \\
04114-5117 & 1.03 & 0.97 & 23230-6926 & 0.58 & 1.50 \\
04384-4848 & 0.93 & 2.08 & 23253-5415 & 0.79 & 1.58 \\
06009-7716 & 1.64 & 3.63 & 23327+2913 & 0.0 & 0.97 \\
06035-7102 & 0.92 & 1.11 & 23365+3604 & 0.0 & 4.47 \\
06206-6315 & 0.33 & 3.69 & 23389-6139 & 0.0 & 1.33 \\
06301-7934 & 0.48 & 1.02 & 23515-2917 & 2.87 & 1.36 \\
06361-6217 & 2.0 & 0.67 & F23529-2119 & 15.32 & 0.38 \\
09104+4109 & 100. & 0.48 & Arp220 & 0.0 & 4.20 \\
12112+0305 & 0.39 & 3.13 & Mrk1014 & 100. & 0.62 \\
14348-1447 & 0.0 & 3.55 & Mrk231 & 100. & 0.31 \\
15250+3609 & 0.46 & 2.88 & Mrk273 & 1.24 & 1.90 \\
15307+3252 & 100. & 0.27 & NGC6240 & 0.0 & 2.58 \\
16474+3430 & 0.35 & 3.58 & UGC5101 & 0.33 & 2.12 \\
16487+5447 & 0.52 & 2.19 & \nodata & \nodata & \nodata \\
\enddata
\tablenotetext{~}{The table includes the ISOPHOT-S \citep{rigopoulou99} and 
CAM-CVF (this paper) ULIRGs.} 
\tablenotetext{a}{Ratio of the integrated
mid-infrared fluxes contained in the continuum contribution and in the
starburst contribution
$F_{5-10{\mu}m}({\rm Continuum})$/$F_{5-10{\mu}m}({\rm Starburst})$ (Cont/Starb,
not dereddened) computed with the model 3a}
\tablenotetext{b}{PAH Line to Continuum ratio}
\end{deluxetable}

\newpage

\begin{table}
\caption{Starburst luminosity to total luminosity ratio}
\begin{tabular}{lrrrr}
\tableline
\tableline
Luminosity &\multicolumn{2}{c}{ $\frac{F_{5-10{\mu}m}({\rm Starformation})} 
{F_{5-10{\mu}m}({\rm Total})}$ }& \multicolumn{2}{c}{
$\frac{F_{3-1000{\mu}m}({\rm Starformation})}{F_{3-1000{\mu}m}({\rm Total})}$} \\
  &  scenario A & scenario B & scenario A & scenario B \\
\tableline
\tableline

$L<10^{12.4}L_{\sun}$ & $65\pm4$  & $84\pm4$   &  $82\pm10$ &   $94 \pm10$   \\

\tableline

$L>10^{12.4}L_{\sun}$ & $24\pm7$ &  $33\pm10$   &  $44\pm17$ &  $55\pm23$ \\

\tableline
\tableline

\end{tabular}
\tablenotetext{~} 
{
Scenario A : continuum is attributed only to AGN ; Scenario B : faint continua
($<$50\% of the total MIR emission) are attributed to a starburst (see  
Section 5). 
The uncertainties are uncertainties of the mean value in one luminosity bin. 
The means of the two luminosity bins of a single scenario differ at 5 sigma in
the MIR.
}
\label{tab:RATAS_Lir}
\end{table}


\newpage

\figcaption[Dist_Lir] 
{Histogram of infrared luminosities for ULIRGs with low resolution ISO
mid-infrared spectroscopy (taken from ISOPHOT-S observations
\cite{rigopoulou99} and ISOCAM-CVF observations (this paper)). Sources
with ISOCAM-CVF spectra shown in this paper are shaded. The
combined sample shown here is biased towards high luminosity ULIRGs.
\label{Dist_Lir}}

\figcaption[ESS_PLOT_ALL2]
{Mid-infrared spectra of sources observed with ISOCAM.  Shaded regions
represent the sum of RMS and systematic error, as described in section
2.2. Positions of the 6.2, 7.7, 8.6 and 11.3 $\mu$m features are
indicated in each plot. We also plot the uncorrected spectrum of IRAS
F02455-2220 which should be considered a lower limit (see text). An
upper limit is given for IRAS 02115+0226. Wavelengths are rest-frame
wavelengths.
\label{results_spectrum}}

\figcaption[COMP_LC_NB_2.PS]
{Comparison between the L/C diagnostic of the 7.7$\mu$m band adopted
by Genzel et al. (1998) and subsequent papers, and a narrow band
diagnostic using just the ratio of fluxes at 7.7 and 6$\mu$m for
Galaxies observed by PHOT-S \citep{rigopoulou99}, SWS \citep{sturm00},
and CAM-CVF (this paper). Upper and lower limits have been suppressed
in this plot. The overplotted line shows the expected relation if the
underlying continuum is flat.
\label{LC_NB}}

\figcaption[EXTINCTION_LAWS.PS]
{Comparison of the extinction laws described in section 4. The Draine
and Lee law is lower than the Galactic center law. The squares
represent the extinction law towards the GC derived from hydrogen
lines
\citep{lutz99a}.
\label{GC_ext}}

\figcaption[disp_fit_sample_new.eps]
{Examples of model fits to ULIRG spectra, using model 3a. The figure
includes high and low S/N spectra, and AGN-dominated and
starburst-dominated spectra to show the success of the fitting
procedure in all those cases. Data are shown by thin lines while the
fitted spectra are shown by thick lines.
\label{disp_fit}}

\figcaption[proof_cena.eps]
{Test of the quality of the two model classes, using mixed
starburst/AGN templates (upper plots: Cen A, lower plots:
Circinus). The left plots show models (3a) where starburst and AGN are
subject to different levels of extinction, while the extinction is
identical for both components in the right plots (model 2). Models
with different extinction for AGN and starburst are clearly favored.
\label{comp_visu}}

\figcaption[stb.eps]
{Comparison of the fitted spectrum and the model (3a) with the
starburst templates M83, NGC3256, NGC 6946 and NGC 253. The thin plain
line is the observed spectrum, the thick line is the model spectrum,
the dashed line is the continuum contribution and the dotted line is
the starburst contribution.
\label{verif_Stb}}

\figcaption[AGN.eps]
{Comparison of the fitted spectrum and the model for the continuum/AGN
templates spectrum : SgrA* (model 3b) and NGC 1275, NGC 1068 and NGC
5506 (model 3a).  The thin plain line is the observed spectrum, the
thick line is the computed spectrum, the dashed line is the continuum
contribution and the dotted line is the starburst contribution (not
visible for GC-SgrA and NGC 1275). The GC SgrA spectrum is not
perfectly reproduced at short wavelengths because of the extinction
law used.
\label{verif_Agn}}

\figcaption[mix.eps]
{Comparison of the fitted spectrum and the model for the mixed
templates : Circinus (observed by SWS, \cite{sturm00}), Centaurus A,
Mrk 273 and NGC 7469.  Graphical conventions are the same than in
Figure~\ref{verif_Agn}. The observed AGN spectra of CenA (Mirabel {\it
et al.} 1999), and Circinus (Moorwood et al. 2000), observed with
ISOCAM-CVF, are overplotted (thick plain line) and compared to the
continuum component deduced from the model. In both cases, the
observed AGN component is similar to the continuum contribution
derived from the model.
\label{verif_mix}}

\figcaption[RATAS_ULIRG_ONLY.PS]
{Comparison between the L/C ratio of the 7.7 $\mu$m band and the
ratio of 5-10$\mu$m Continuum and starburst UIB spectrum
($F_{5-10{\mu}m}({\rm Continuum})$/$F_{5-10{\mu}m}({\rm Starburst})$)
given by the fit 3a (not dereddened). Squares represents ULIRG points,
diamonds starbursts, and triangle AGNs. A 5-10$\mu$m continuum to
starburst flux ratio greater than 50 is representative of a spectrum
dominated by the continuum component. In order to maintain readability
of the figure we assigned to those galaxies arbitrary continuum to
band flux ratio near 100.
\label{LC_RATAS}}

\figcaption[LC_DIETER.PS] {Line to continuum ratio of the 7.7 $\mu$m
  feature as a function of the IR luminosity. The little asterix and
  upper limits are from the sample of ULIRG observed by PHOT-S
  \citep{lutz98,rigopoulou99}. Observations of hyperluminous galaxies
  \citep{taniguchi97,aussel98} are added.
\label{LC_comp}}

\figcaption[arp220_knotA_comp.eps]
{Comparison of the spectra of Arp220 and the strong star forming
region in the overlap region of NGC4038/4039 (KnotA , Vigroux et. al
(1996), Mirabel et al. (1998)). The graphical convention are as 
described for Figure~\ref{verif_Agn}.
\label{arp220_knotA_comp}}

\figcaption[ratas_lir_med_17_fb.ps] { Trend of the mid-infrared ratio
  $F_{5-10{\mu}m}({\rm Starburst})$/$F_{5-10{\mu}m}({\rm Total})$ (not
  dereddened) versus infrared luminosity.  At each luminosity, the
  plot shows the median $F_{5-10{\mu}m}({\rm
    Starburst})$/$F_{5-10{\mu}m}({\rm Total})$ of the 17 galaxies from
  Table~\ref{tab:ulirg_ratas_lc} and \citep{taniguchi97,aussel98}
  closest to the considered luminosity. The right axis displays the
  extrapolated bolometric ratio $F_{3-1000{\mu}m}({\rm
    Starformation})$/$F_{3-1000{\mu}m}({\rm Total})$ as described in
  section \ref{res_mod}.\label{Lir_mirfrac}}

\newpage
\plotone{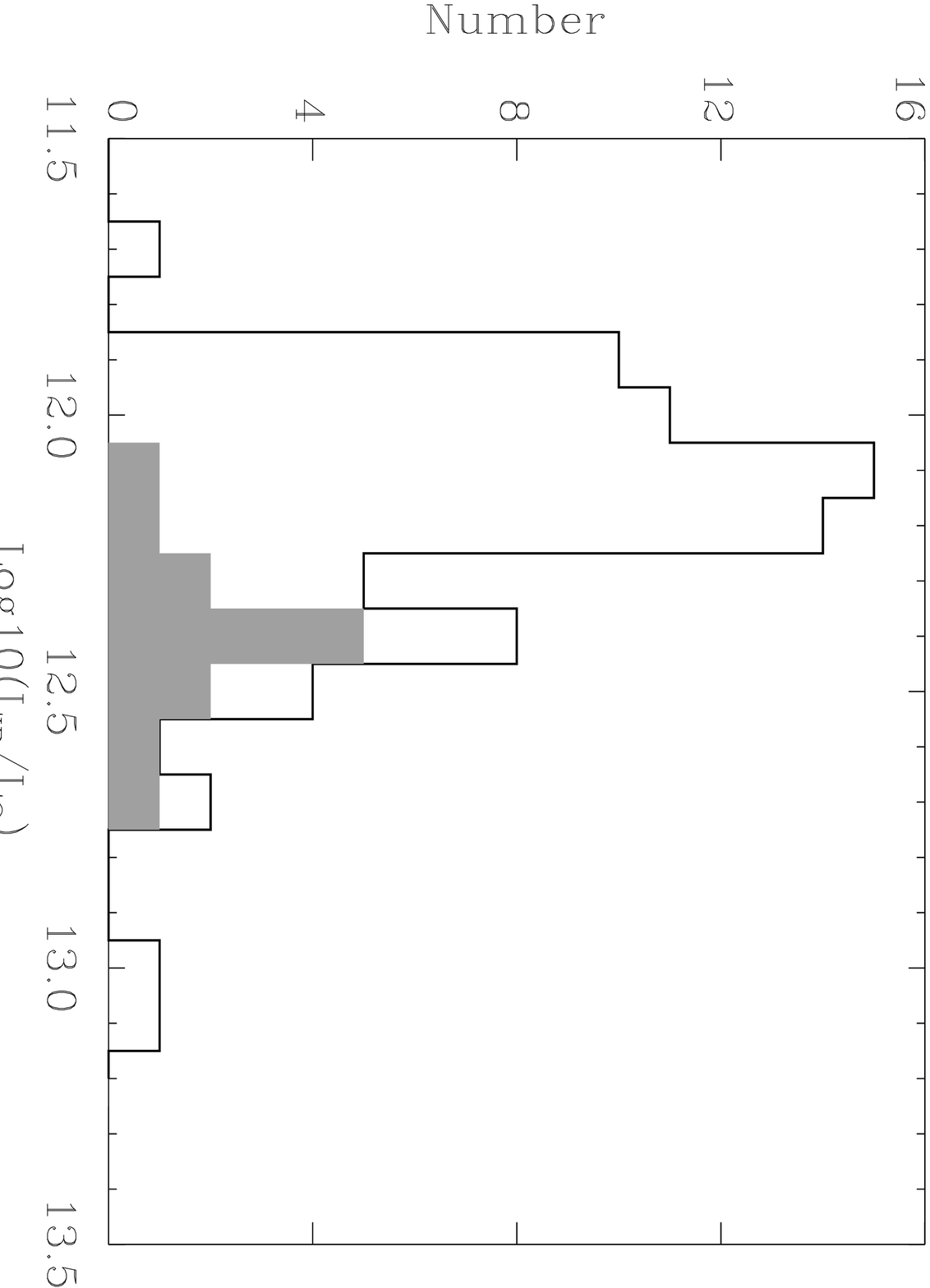}

\newpage
\plotone{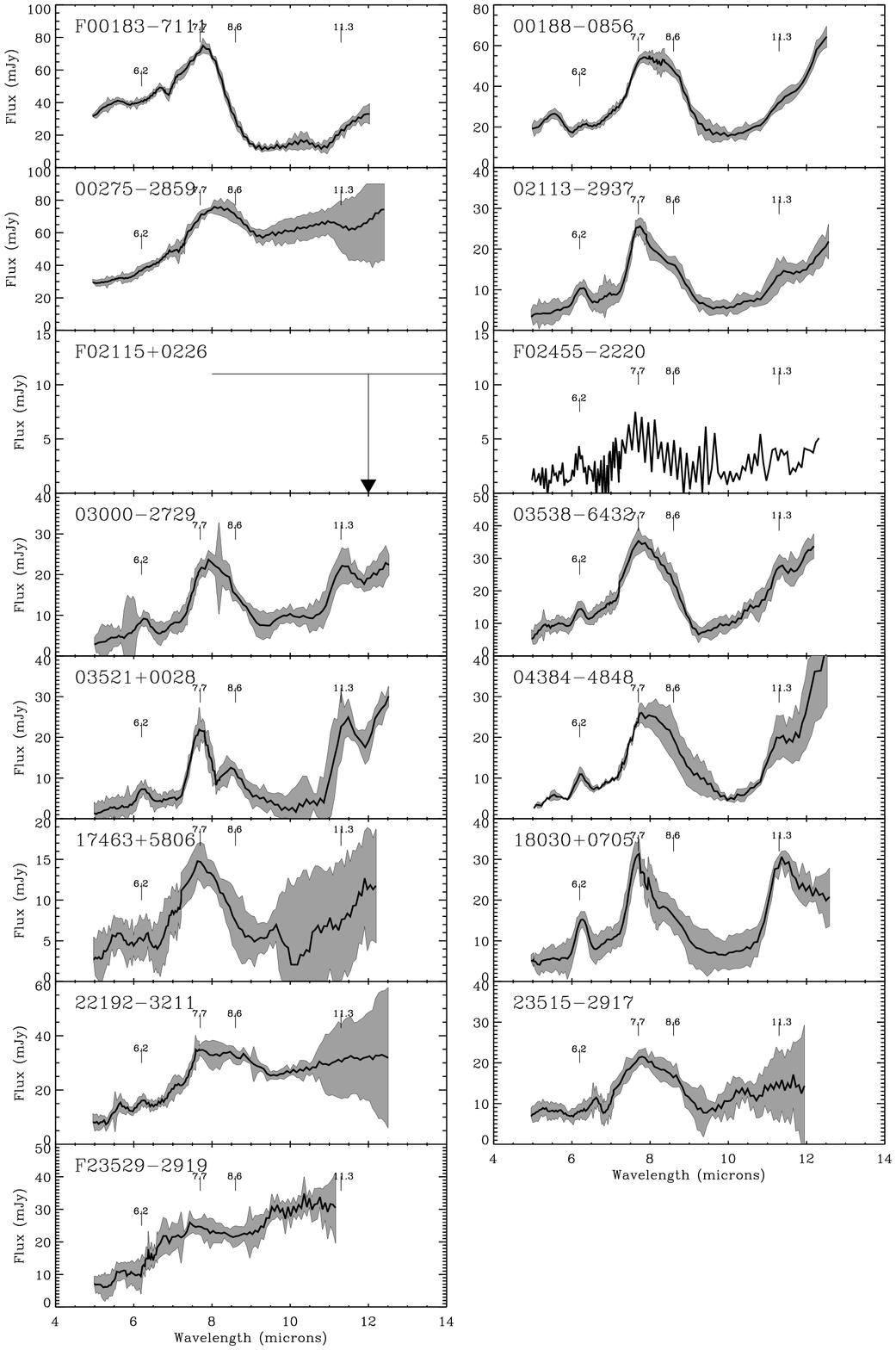}

\newpage
\plotone{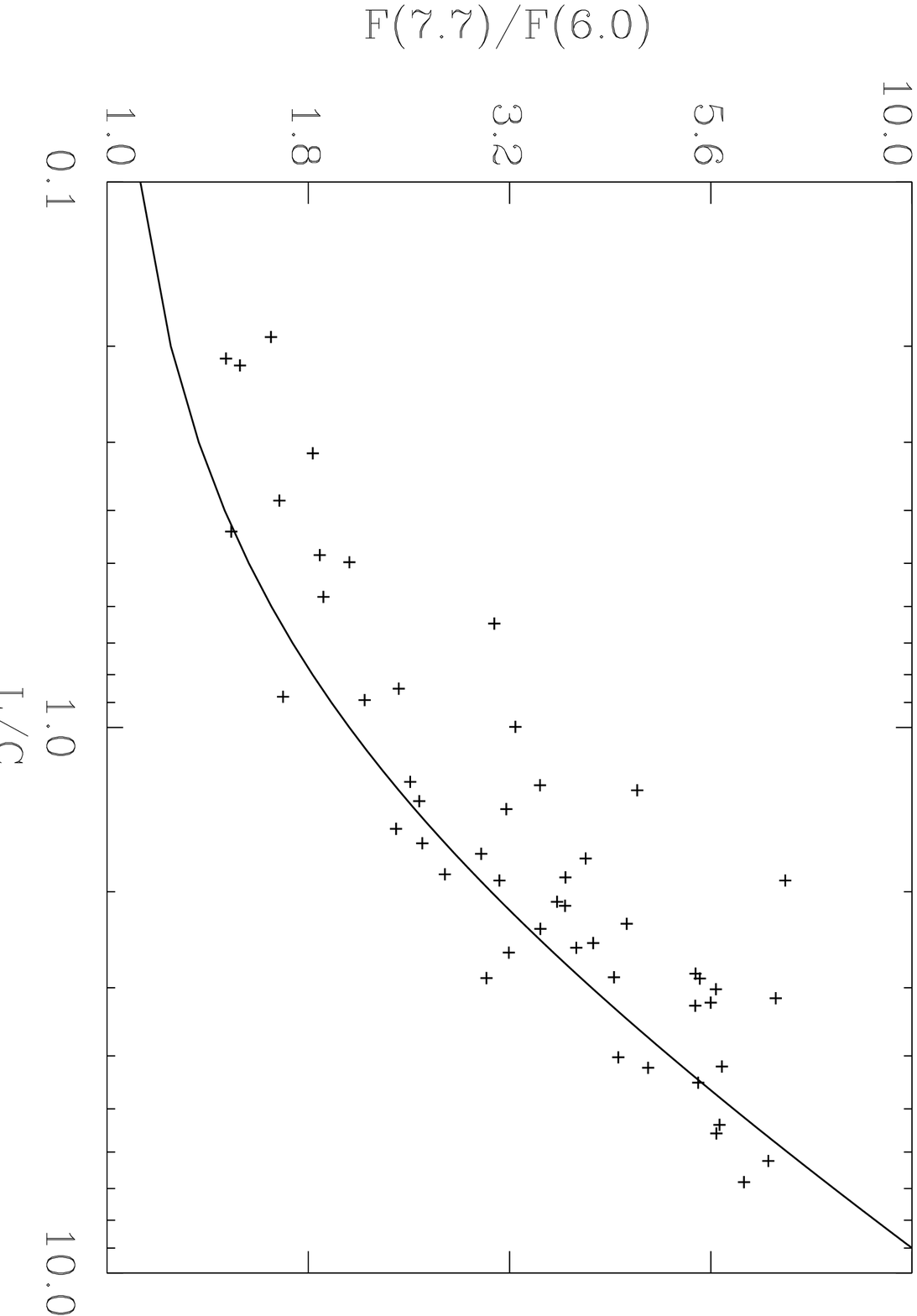}

\newpage
\plotone{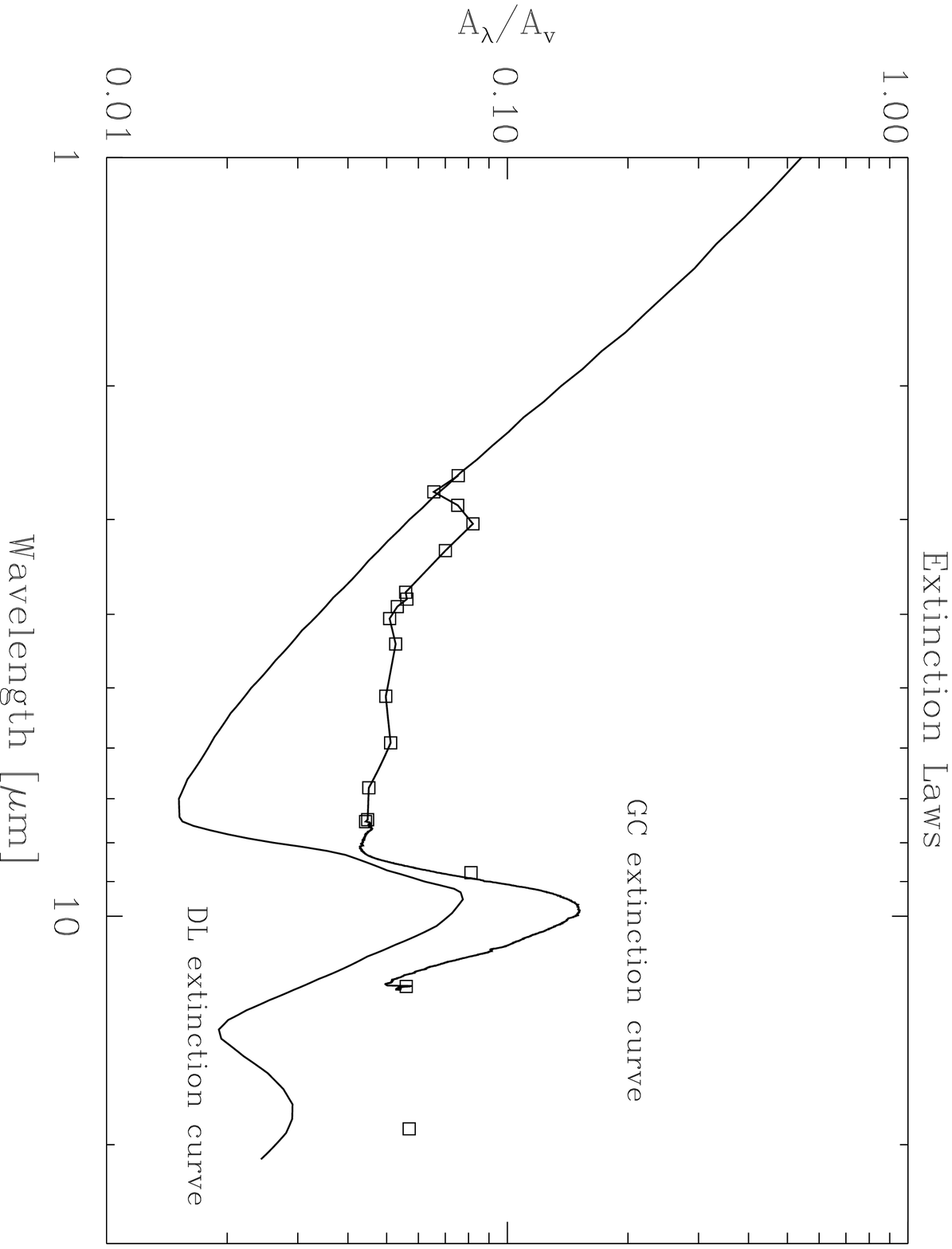}

\newpage
\plotone{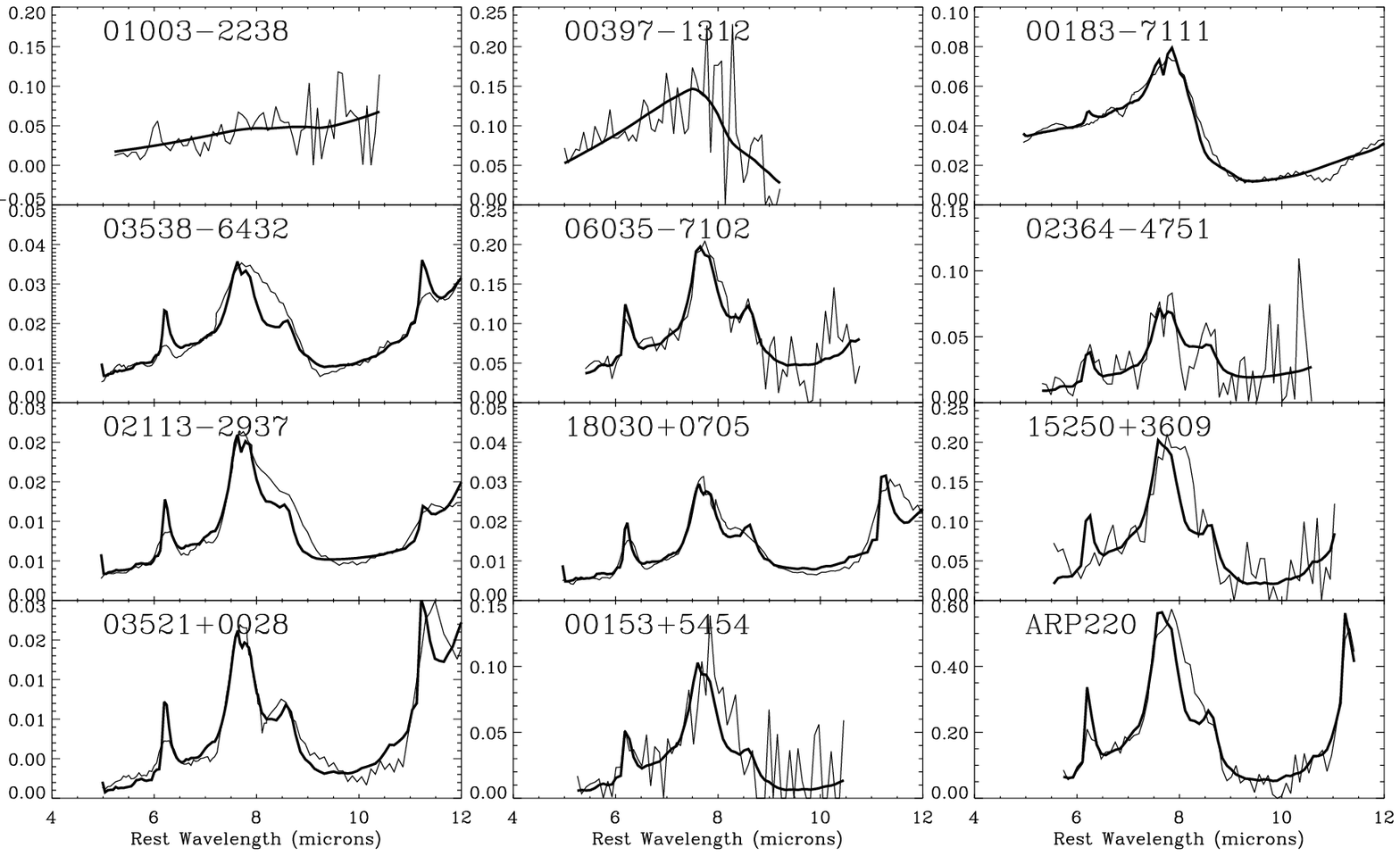}

\newpage
\plotone{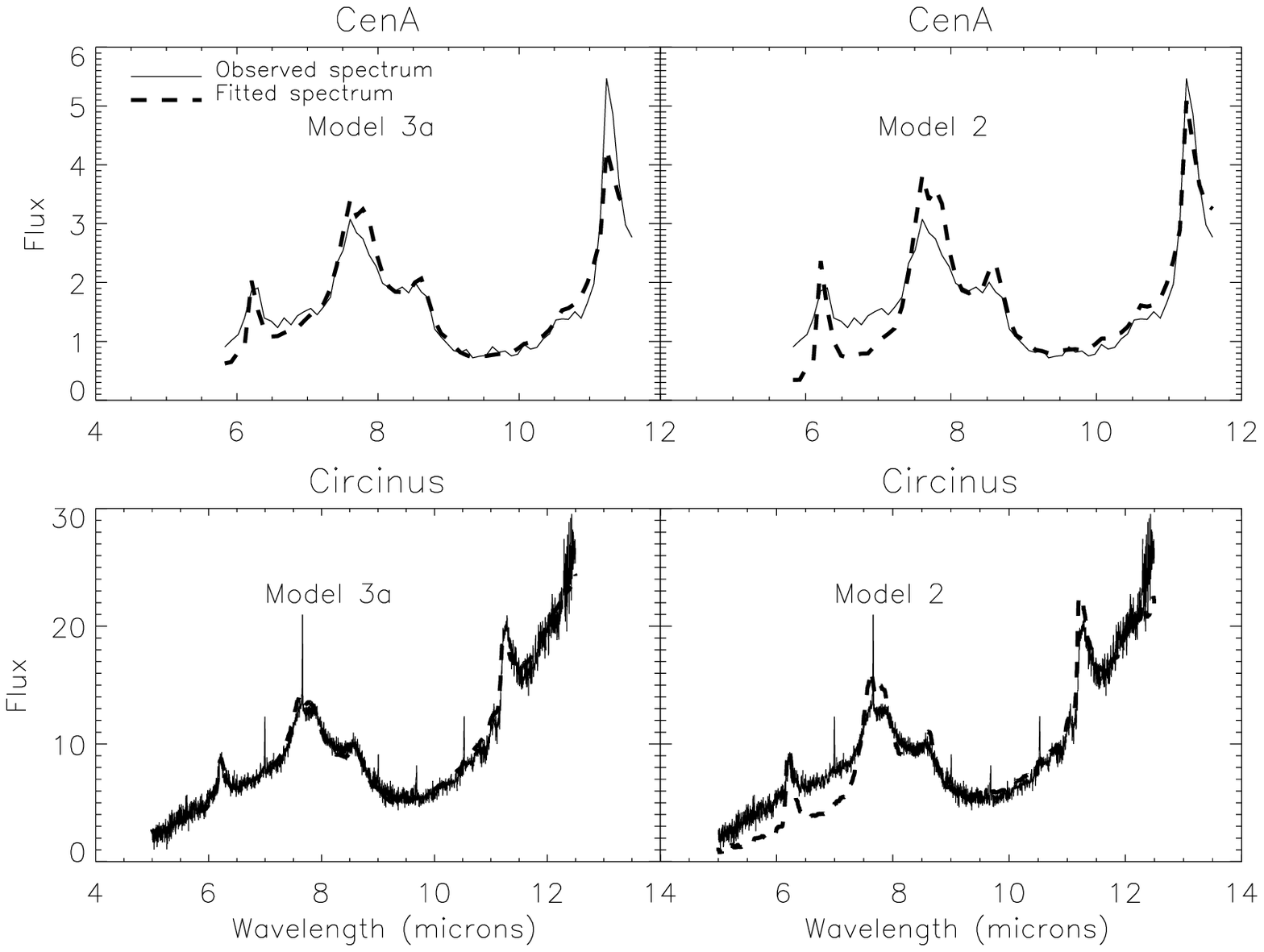}

\newpage
\plotone{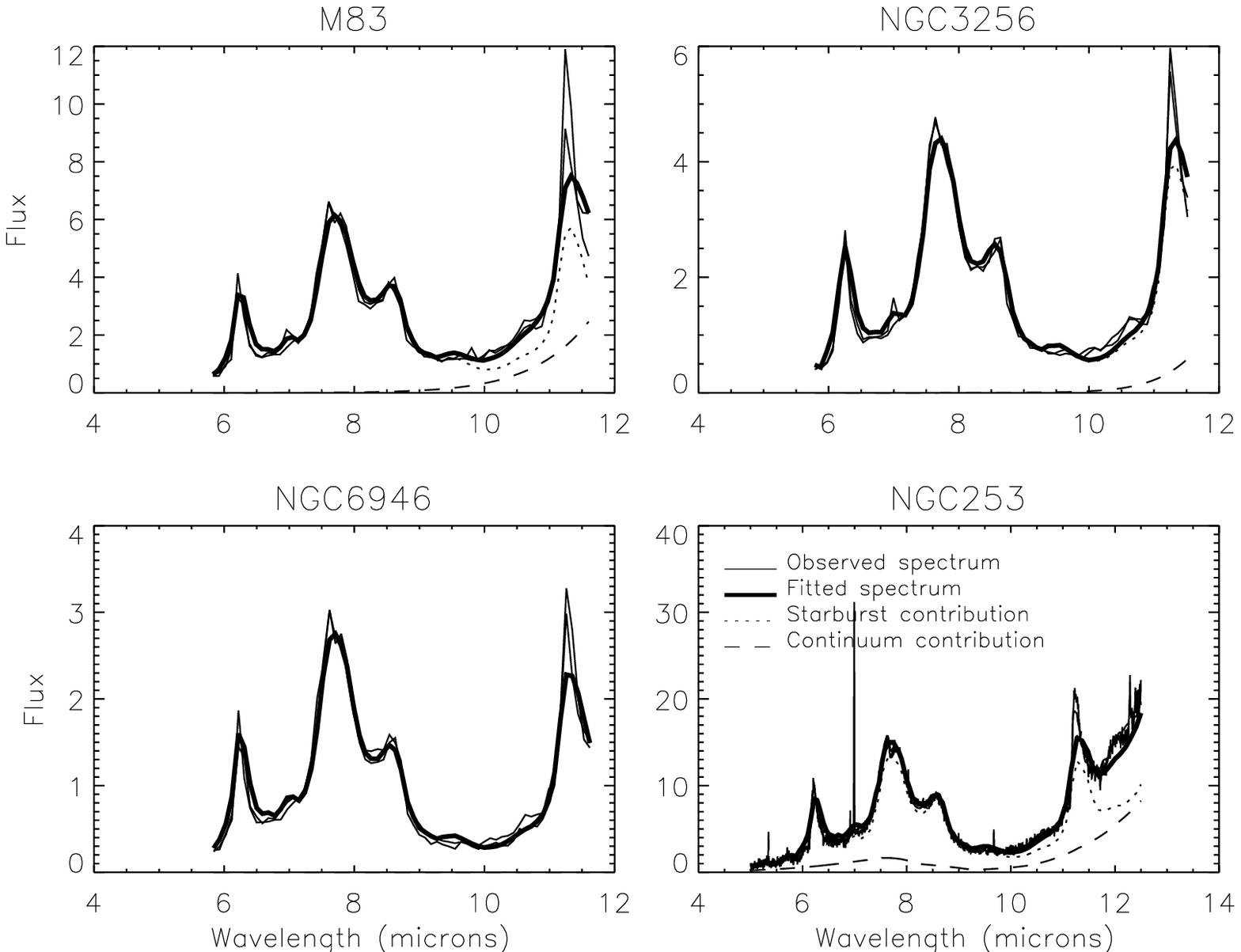}

\newpage
\plotone{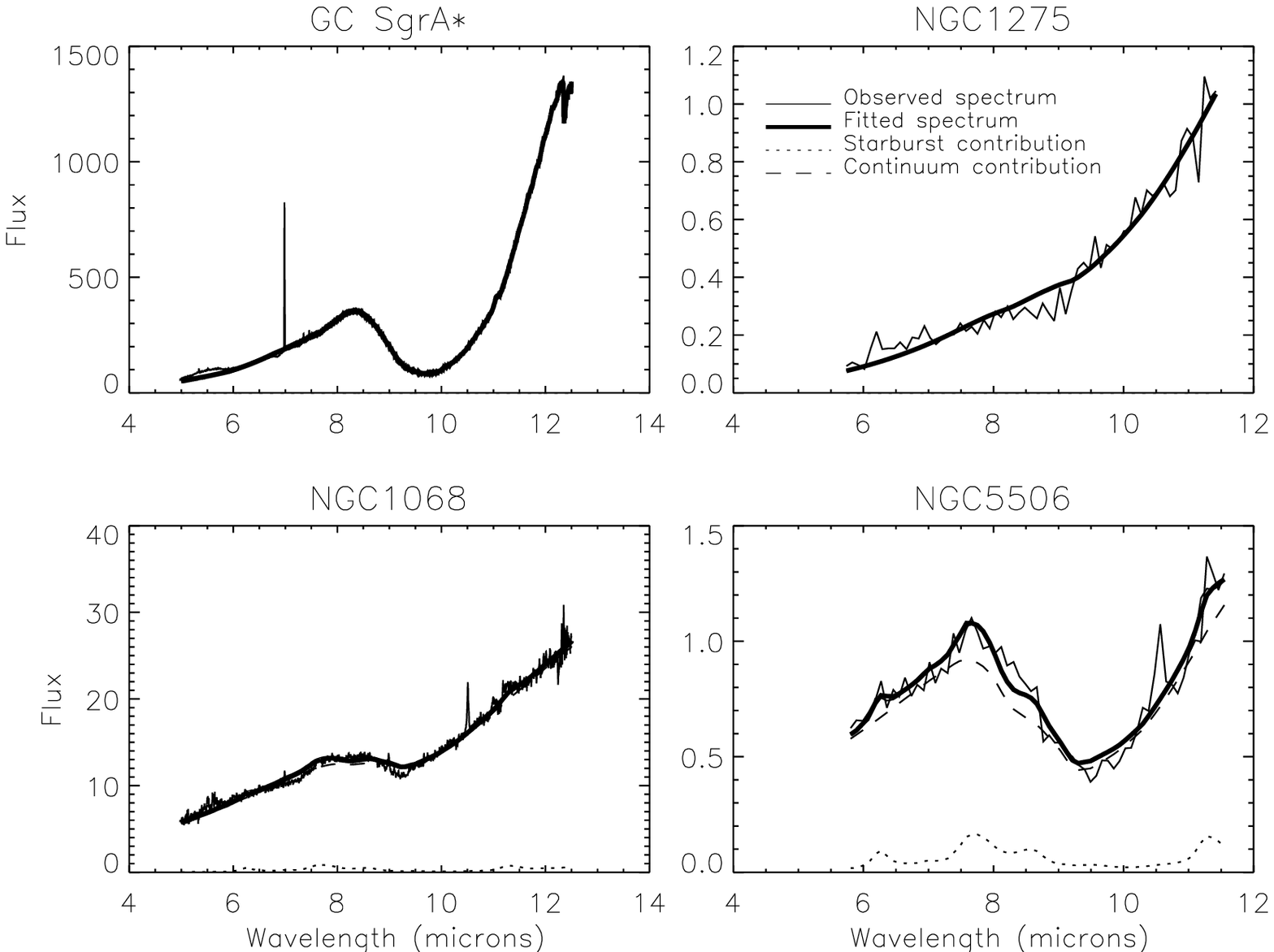}

\newpage
\plotone{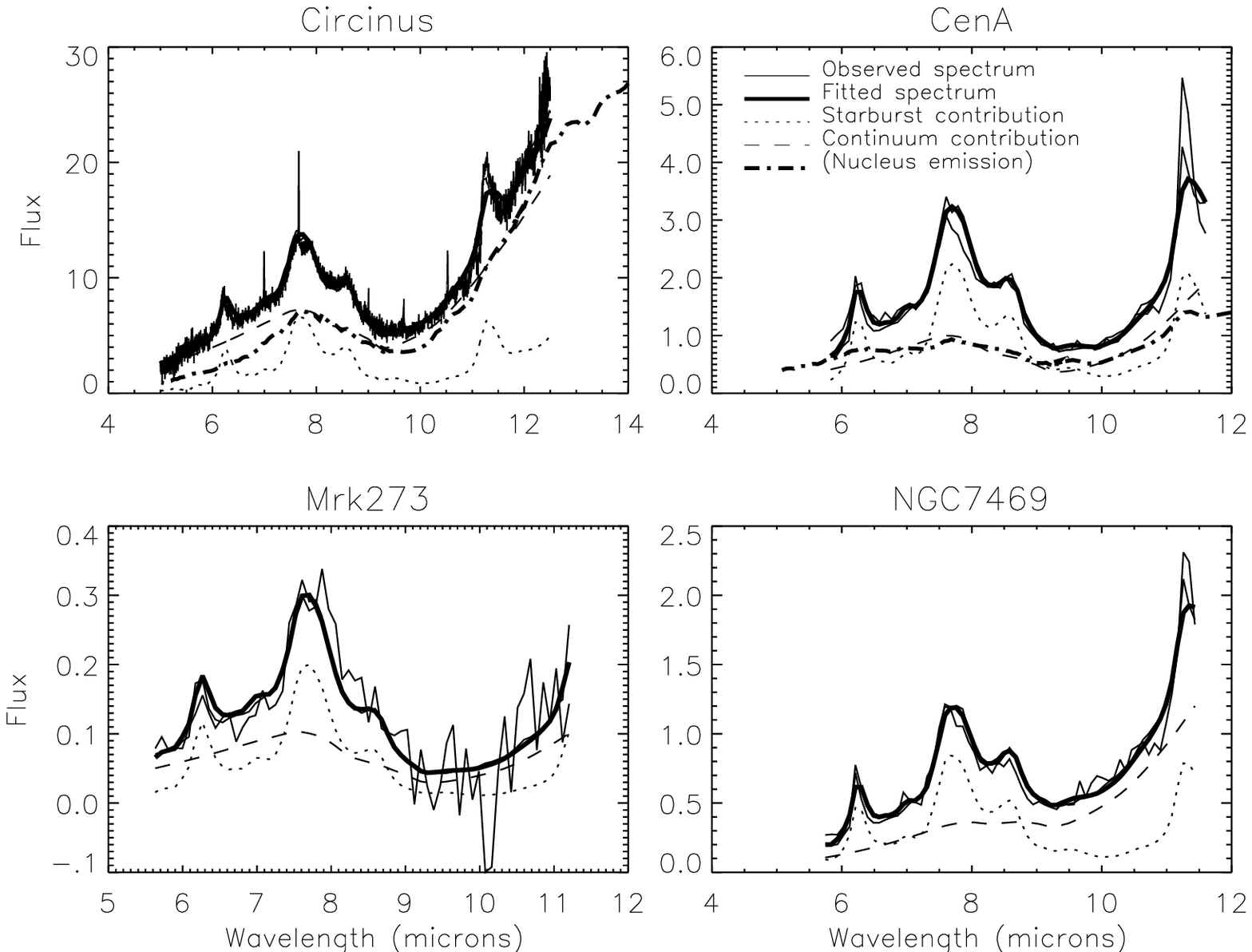}

\newpage
\plotone{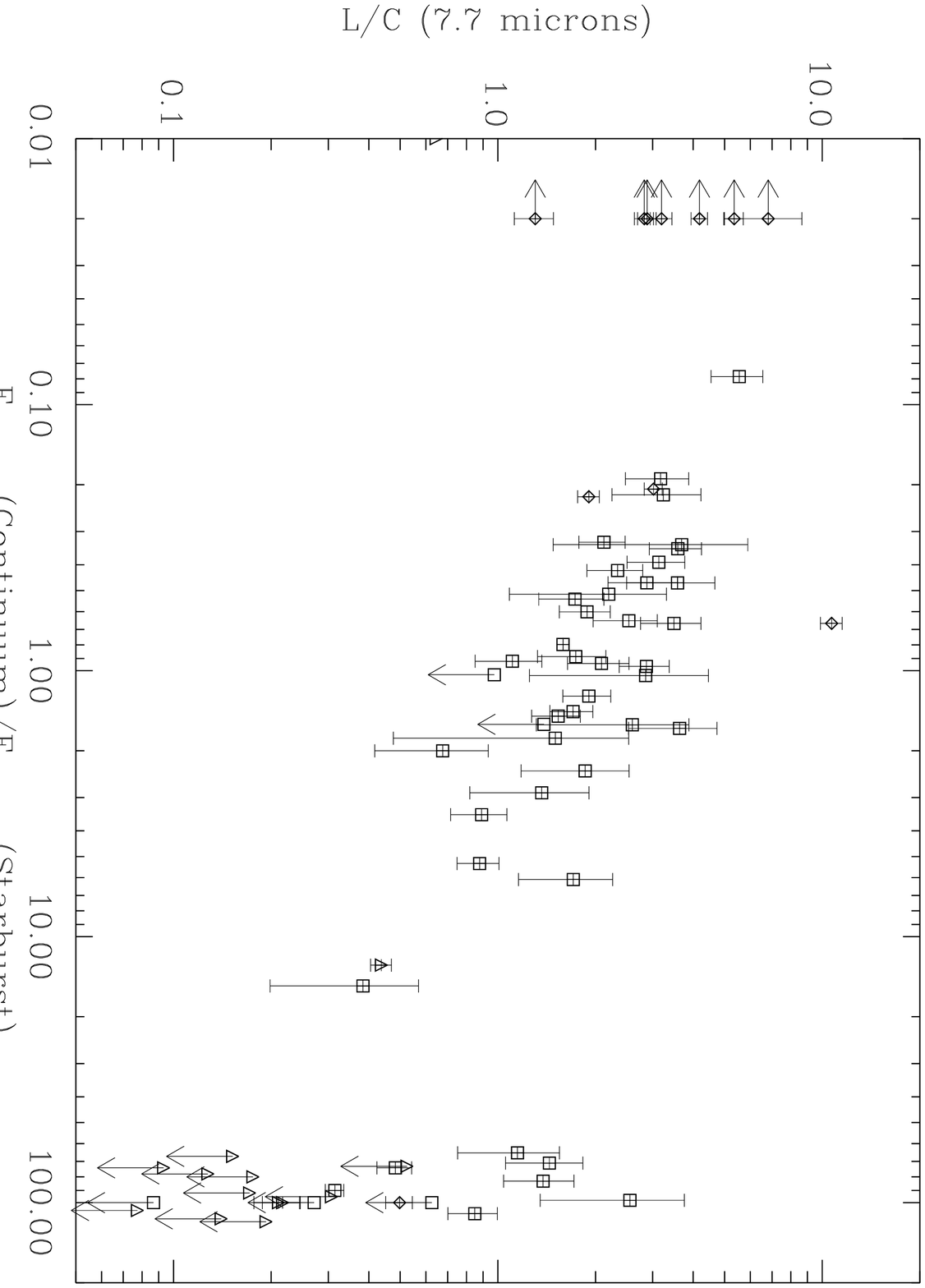}

\newpage
\plotone{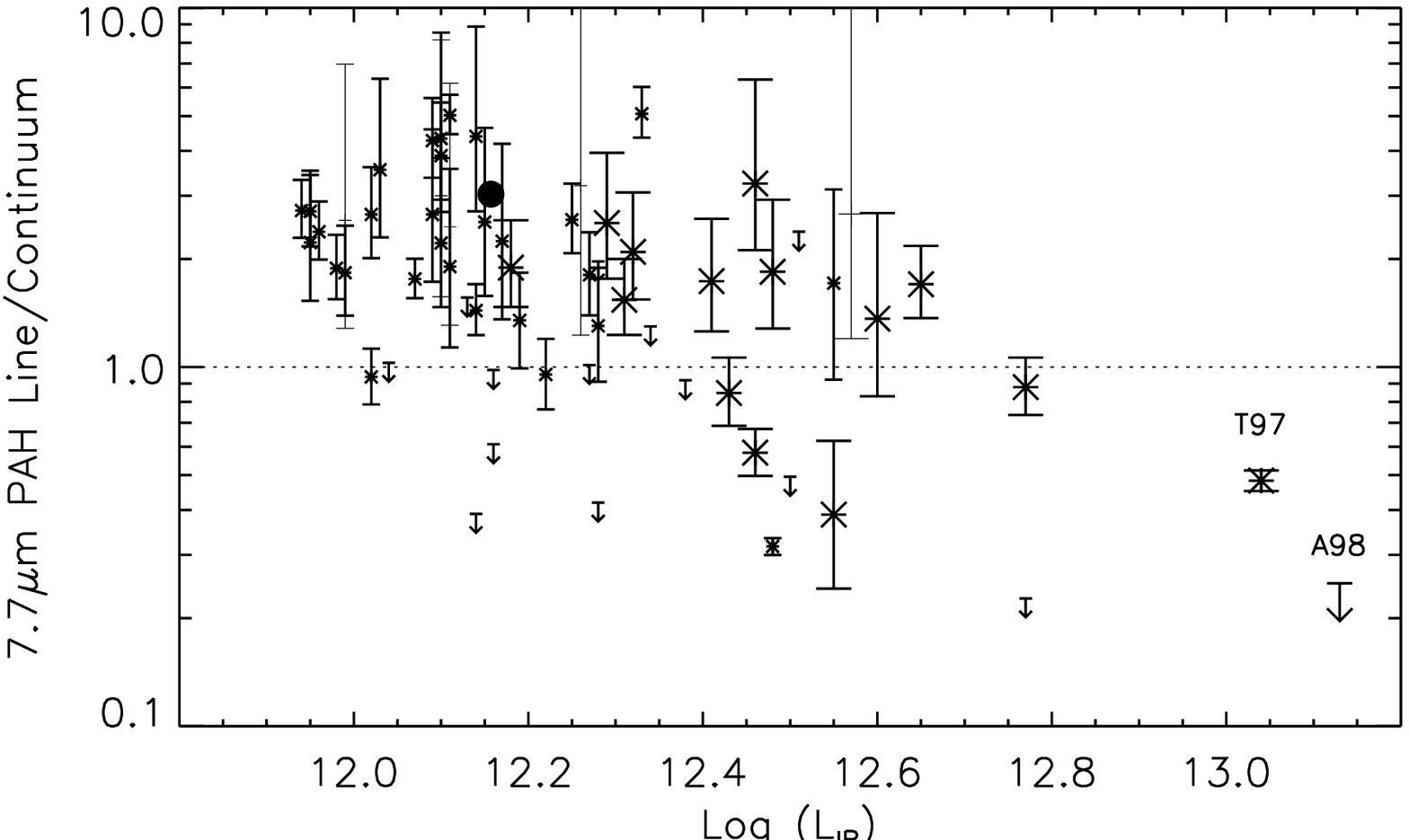}

\newpage
\plotone{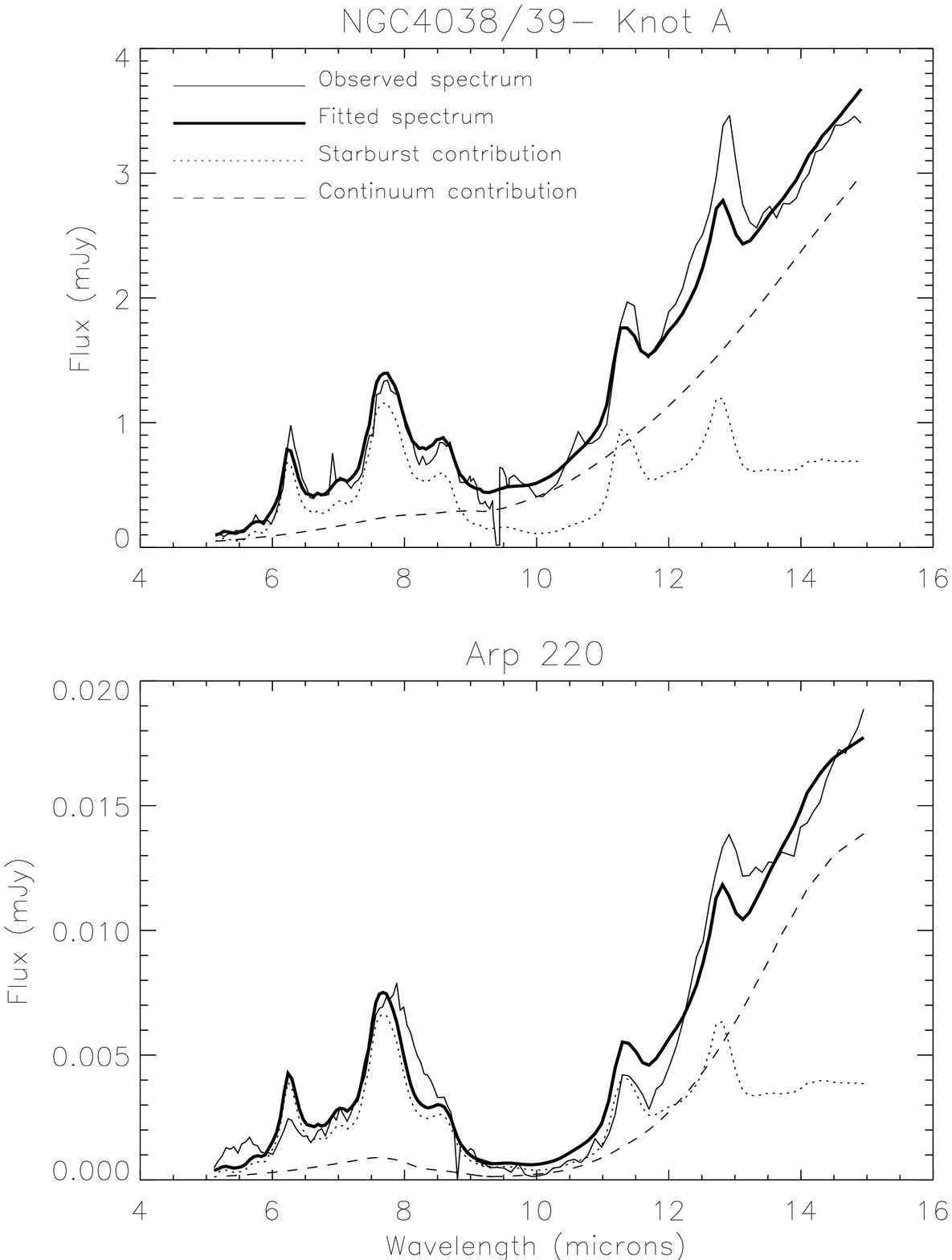}

\newpage
\plotone{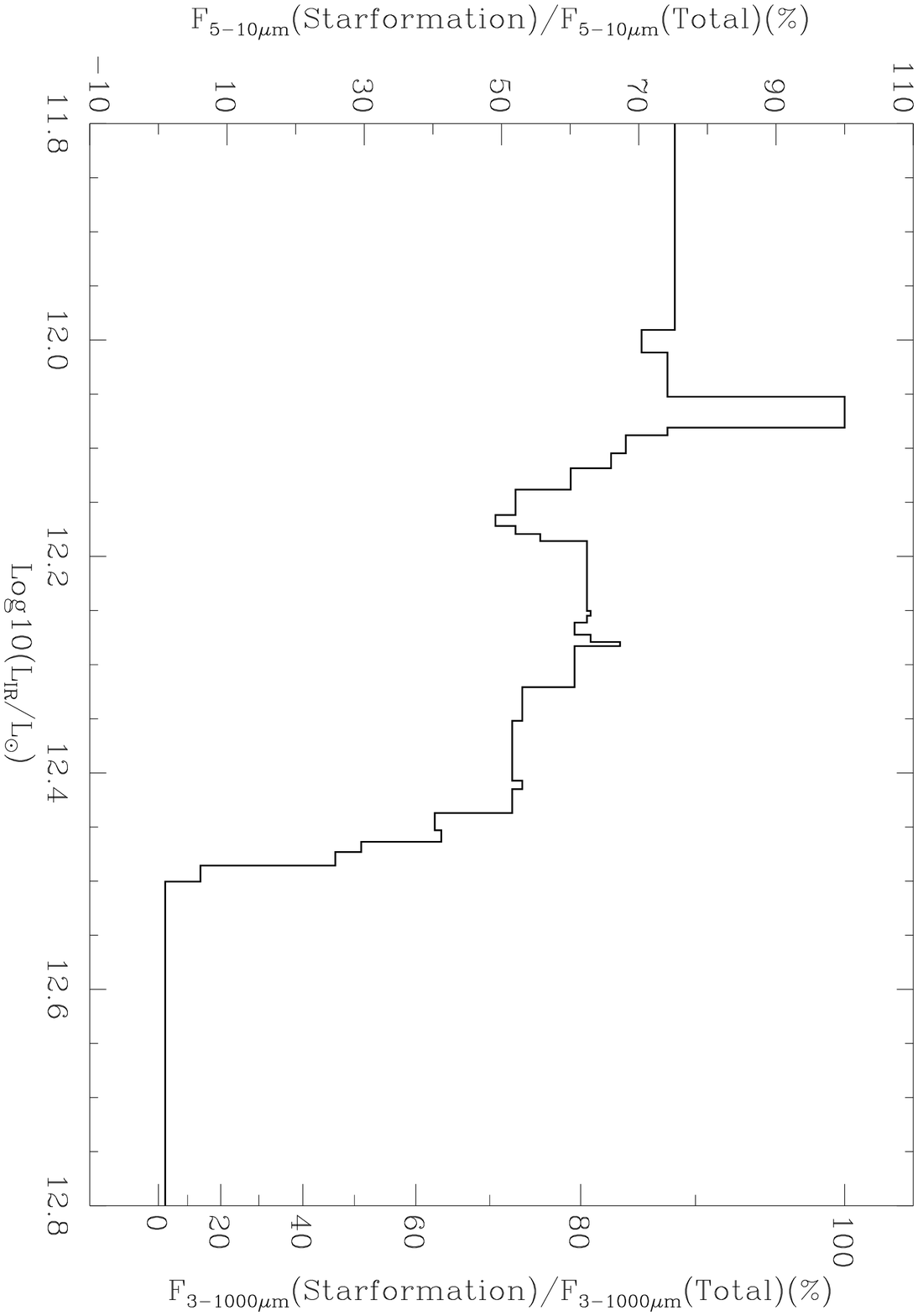}

\end{document}